\documentclass[preprint,12pt,number]{elsarticle}

\usepackage{float}

\usepackage{caption}
\usepackage{amsmath,amssymb,amsfonts}
\usepackage{booktabs}
\usepackage{multirow}
\usepackage{array}
\usepackage{graphicx}
\usepackage{enumitem}
\usepackage{xcolor}
\usepackage{tikz}
\usepackage{rotating}
\usepackage{hyperref}
\usepackage{tabularx}
\usepackage{booktabs}
\usepackage{pifont} %
\usetikzlibrary{arrows.meta,positioning,fit,calc,shapes.geometric}

\journal{Aerospace Science and Technology}
\graphicspath{{fig/}}
\hypersetup{colorlinks=true,linkcolor=blue,citecolor=blue,urlcolor=blue}
\setlist[itemize]{leftmargin=1.8em}
\setlist[enumerate]{leftmargin=1.8em}

\newcommand{\ood}{\textit{OOD}}
\newcommand{\id}{\textit{ID}}

\newcommand{\rmse}{\textit{RMSE}}
\newcommand{\mae}{\textit{MAE}}
\newcommand{\sharedparam}{\textit{TFCN-SP}}
\newcommand{\typedparam}{\textit{TFCN-TP}}

\begin{document}

\begin{frontmatter}

\title{Compositional Embedding Architecture for Physical Field Prediction in Componentized Aerospace Systems}

\author[inst1]{Qineng Wang}
\ead{wangqn@microsate.com}

\author[inst1,inst2]{Xinrui Zhou}

\author[inst1]{Shuwen Yue}

\author[inst1]{Kangli Bao}

\author[inst1]{Hairun Xie\corref{cor1}}
\ead{xiehr@microsate.com}

\author[inst1]{Yonghe Zhang\corref{cor1}}
\ead{zhangyh@microsate.com}

\cortext[cor1]{Corresponding authors.}
\affiliation[inst1]{
  organization={Key Laboratory for Satellite Digitalization Technology, Innovation Academy for Microsatellites, Chinese Academy of Sciences},
  city={Shanghai},
  country={China}
}
\affiliation[inst2]{
  organization={School of Aeronautics and Astronautics, Shanghai Jiao Tong University},
  city={Shanghai},
  country={China}
}
\begin{abstract}
Spacecraft thermal design requires repeated evaluation of how variations in the number and spatial arrangement of heat-generating components and in thermal boundary conditions affect the temperature field. High-fidelity numerical simulations are computationally expensive and therefore difficult to use for large-scale design screening. Although surrogate models can accelerate temperature-field prediction, existing approaches generally encode each complete configuration as a whole and do not explicitly exploit the reusability of local physical constituents across configurations, which limits their accuracy for component counts and combinations not covered during training. To address this issue, we propose the Tree-Structured Factor Composition Network (TFCN), which decomposes complex spacecraft thermal configurations into reusable local physical factors and employs a tree-structured composition module to learn the global temperature-field response associated with different factor combinations.

TFCN is evaluated on two-dimensional steady-state spacecraft thermal-analysis cases with prescribed-temperature and radiative-flux boundary conditions. The model is trained exclusively on configurations containing no more than 15 heat-generating components and evaluated on unseen configurations containing 16–25 components. For the prescribed-temperature and radiative-flux cases, TFCN achieves component-count out-of-distribution RMSE values of 4.21 K and 18.62 K, respectively, representing reductions of 65.6\% and 33.6\% relative to the strongest baseline. These results demonstrate that TFCN improves the reliability of temperature-field prediction under variations in component count and provides an efficient surrogate for rapid spacecraft thermal-design evaluation and large-scale configuration screening.

\end{abstract}


\end{frontmatter}

\section{Introduction}

The design of complex aerospace systems involves continuous configuration and iteration of structures, payloads, equipment units, and functional components under multidisciplinary constraints. Spacecraft are representative componentized aerospace systems composed of platform structures, payloads, electronic equipment, and thermal-control components. As mission requirements and design stages evolve, the types, numbers, layouts, parameters, and operating states of these components may change. Each design modification may alter the loads, heat sources, and boundary conditions of the system, thereby changing key physical responses such as temperature and stress fields. Rapid and reliable prediction of the physical fields associated with candidate configurations is therefore fundamental to design comparison, risk identification, and engineering decision-making. In spacecraft thermal design specifically, onboard equipment and payloads must be maintained within their allowable temperature ranges under the space thermal environment and different operating modes \citep{lv2024review}. Rapid evaluation of the full-field temperature distribution, maximum temperature, and hot-spot location facilitates the identification of potential thermal risks and the comparison of alternative equipment layouts \citep{zhang2025layout}. It also supports the assessment of candidate thermal-control schemes during configuration design \citep{huo2025layout}.

High-fidelity numerical simulations remain essential to spacecraft thermal analysis and final engineering verification. However, each change in equipment layout, power dissipation, or thermal boundary conditions requires repeated model updates, numerical solution, and result post-processing, making such simulations impractical for rapidly screening many candidate configurations. Surrogate models therefore provide low-cost approximations for early design comparison while leaving detailed verification to high-fidelity analysis. They are now widely used in spacecraft thermal design. Multi-fidelity radial basis function (RBF) neural networks accelerate thermal-design sensitivity analysis \citep{xiong2021sensitivity}, while deep transfer learning supports spacecraft thermophysical surrogates across operating conditions \citep{xiong2022surrogate}. Artificial neural networks (ANNs) predict lunar-orbiter test temperatures \citep{jang2024lunarann}, and proper orthogonal decomposition-based physics-informed machine learning (POD-PIML) predicts system-wide temperature fields \citep{tanaka2023pod}. Dimensionally consistent self-supervision (DSSL) addresses full-field prediction with limited labels \citep{wang2026dssl}; recurrent neural networks (RNNs) model transient spacecraft thermal responses \citep{yamashita2026podrnn}; and graph neural networks (GNNs) estimate temperatures in large spacecraft systems \citep{yamashita2026gnn}. Despite this progress, most surrogates encode each complete configuration as a fixed global input and learn a direct mapping to its field response. Finite simulation datasets cover only a subset of component counts, layouts, operating states, and boundary organizations. Consequently, strong in-distribution accuracy does not ensure reliable prediction for unseen component counts or new combinations of heat sources and boundaries. This limitation makes out-of-distribution generalization a central requirement for componentized spacecraft thermal surrogates and motivates examining inductive structures that support extrapolation beyond the configurations represented during training.

Out-of-distribution generalization in componentized systems depends on more than prediction accuracy for configurations observed during training. A model must also accommodate changes in the number and organization of physical objects. This is a general problem in learning on sets, graphs, irregular geometries, and parametrized PDEs, and has motivated extensive studies on variable-size representations and compositional architectures. For the present task, two capabilities are especially relevant: accepting variable-count structured inputs and organizing them into a system-level response. Table~\ref{tab:method_survey} therefore organizes related methods by where their principal generalization bias acts: on physical consistency, spatial or set representation, or mechanism/operator composition.

\begin{table}[htbp]
\caption{Literature-level feature matrix for related physical-field surrogate methods.}
\label{tab:method_survey}
\centering
\scriptsize
\resizebox{\linewidth}{!}{
\begin{tabular}{p{0.12\linewidth}|p{0.16\linewidth}cccc}
\toprule
Category &
Method &
\shortstack{Phys. loss\\constraint} &
\shortstack{Variable-count\\feature input} &
\shortstack{Object/subblock\\encoding} &
\shortstack{Shared recursive\\composition} \\
\midrule
\multirow{4}{*}{\shortstack[l]{Physical\\consistency\\constraints}}
& PINN (2019)\citep{raissi2019pinns} & \(\checkmark\) & \(\times\) & \(\times\) & \(\times\) \\
& PI-ML (2023)\citep{tanaka2023pod} & \(\checkmark\) & \(\times\) & \(\times\) & \(\times\) \\
& PI-DeepONet (2024)\citep{lu2024mbpideeponet} & \(\checkmark\) & \(\times\) & \(\times\) & \(\times\) \\
& DSSL (2026)\citep{wang2026dssl} & \(\checkmark\) & \(\times\) & \(\times\) & \(\times\) \\
\midrule
\multirow{4}{*}{\shortstack[l]{Spatial/set\\representation}}
& DeepSets (2017)\citep{zaheer2017deep} & \(\times\) & \(\checkmark\) & \(\checkmark\) & \(\times\) \\
& GINO (2023)\citep{li2023gino} & \(\times\) & \(\checkmark\) & \(\checkmark\) & \(\times\) \\
& BENO (2024)\citep{wang2024beno} & \(\times\) & \(\checkmark\) & \(\checkmark\) & \(\times\) \\
& GNN (2026)\citep{yamashita2026gnn} & \(\times\) & \(\checkmark\) & \(\checkmark\) & \(\times\) \\
\midrule
\multirow{4}{*}{\shortstack[l]{Mechanism \\ /operator \\ composition}}
& PDE-Net (2019)\citep{long2018pdenet,long2019pdenet2} & \(\checkmark\) & \(\times\) & \(\times\) & \(\checkmark\) \\
& SDNO (2024)\citep{wang2024sdno} & \(\times\) & \(\checkmark\) & \(\checkmark\) & \(\checkmark\) \\
& CompNO (2026)\citep{hmida2026compno} & \(\times\) & \(\times\) & \(\times\) & \(\checkmark\) \\
& LegONet (2026)\citep{zhang2026legonet} & \(\times\) & \(\times\) & \(\checkmark\) & \(\checkmark\) \\
\midrule
\shortstack[l]{Input-factor\\composition}
& \textbf{TFCN (ours)} & \(\times\) & \(\checkmark\) & \(\checkmark\) & \(\boldsymbol{\checkmark}\) \\
\bottomrule
\end{tabular}}
\end{table}

Physical-consistency methods constrain the admissible solution through governing equations, boundary laws, or dimensional relations \citep{tanaka2023pod,wang2026dssl,raissi2019pinns,lu2024mbpideeponet}. Spatial and set representation methods preserve variable input structures through permutation-invariant aggregation, geometry encoding, boundary embedding, or graph representation \citep{yamashita2026gnn,zaheer2017deep,li2023gino,wang2024beno}. Mechanism/operator approaches instead reuse response fragments, differential mechanisms, or learned operator blocks \citep{long2018pdenet,long2019pdenet2,wang2024sdno,hmida2026compno,zhang2026legonet}. Existing methods primarily address variable-size representation or the reuse of equation-level structures.

To address this gap, this study proposes the Tree-Structured Factor Composition Network (TFCN). Each heat-source component or boundary segment is represented as a factor defined by its spatial support and physical scalars. Balanced bucketing and factor encoders convert a variable set of factors into tokens, and a shared tree composer recursively merges these tokens before full-field decoding. TFCN represents an unseen configuration by recursively composing geometry-bearing factor tokens. This distinguishes it from spatial and set methods that aggregate variable-size inputs and from mechanism/operator methods that compose equation-level blocks.

The contributions of this study are threefold.

\noindent\textbf{(i)} We propose TFCN, which performs shared recursive composition over geometry-bearing engineering objects by repeatedly applying a parameter-shared composer to construct system-level latent representations.

\noindent\textbf{(ii)} We generalize engineering composition beyond physical components by incorporating boundary segments and boundary-condition parameters as composable factors. Performing composition in latent rather than physical space provides the framework with potential applicability to other modular PDE problems.

\noindent\textbf{(iii)} Evaluations under Prescribed-Temperature BC and Radiative-Flux BC scenarios demonstrate that TFCN substantially improves high-component-count \ood\ prediction over UNet, DeepONet, and GINO. Structural ablations and engineering thermal metrics further verify the effectiveness of the proposed shared recursive composition mechanism.

The remainder of this paper is organized as follows. Section~\ref{sec:problem_setup} defines the componentized thermal-field prediction problem, including its boundary physics, data construction, and evaluation protocol; Section~\ref{sec:method} presents the TFCN architecture and training strategy; Section~\ref{sec:experiments} reports the main results, ablations, and diagnostic analyses; and Section~\ref{sec:conclusion} concludes the paper and discusses its limitations.

\section{Problem setup}
\label{sec:problem_setup}

\begin{figure}[H]
\centering
\includegraphics[width=0.92\linewidth]{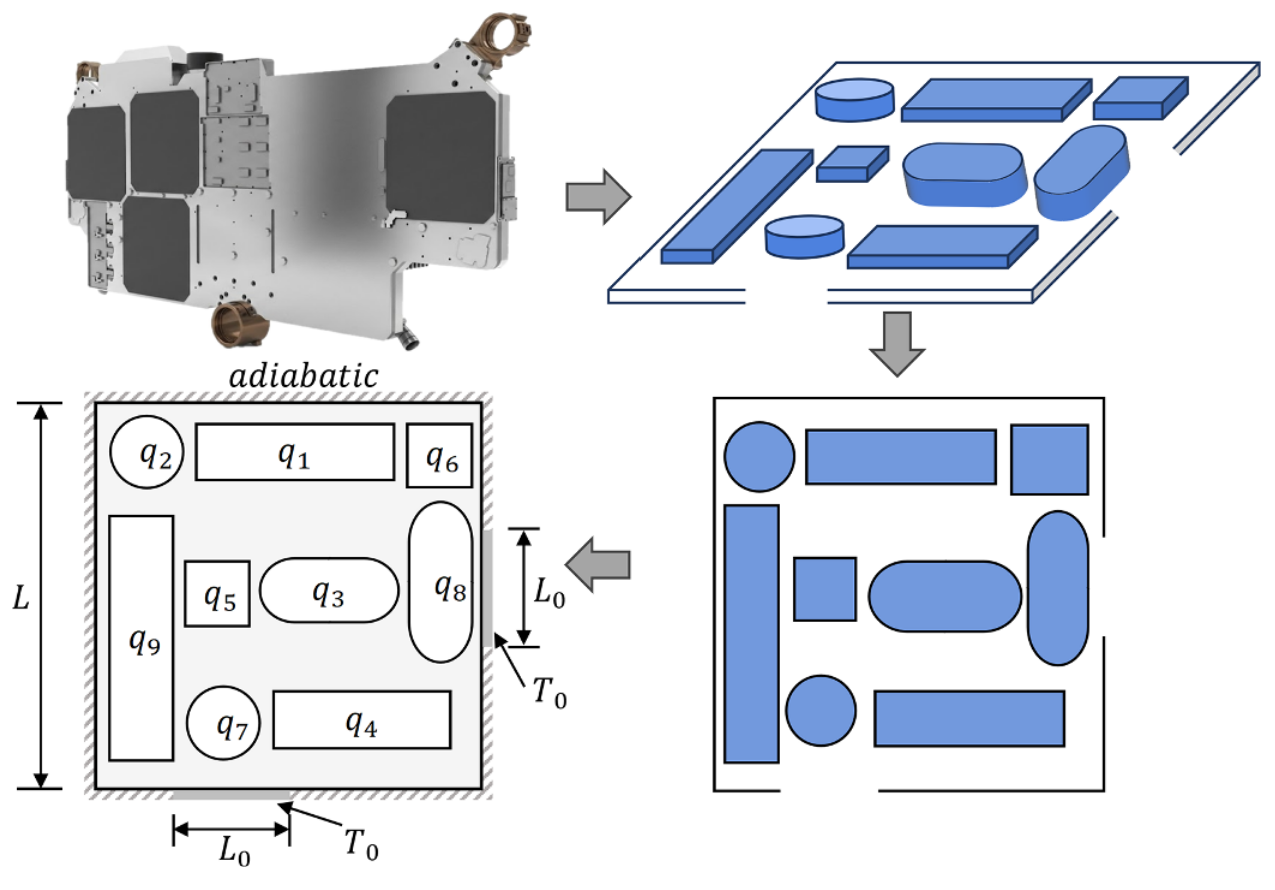}
\caption{From a componentized spacecraft layout to the two-dimensional HSL-TFP problem. The underlying problem follows the general heat-source-layout benchmark of Chen et al.~\citep{chen2021heatlayout}, and its spacecraft-oriented extension follows our previous work \citep{wang2026dssl}.}
\label{fig:problem_abstraction}
\end{figure}

Spacecraft are componentized aerospace systems. Electronic units and payloads occupy different spatial supports and dissipate heat, while their composition, layout, operating states, and heat-rejection conditions determine the structural temperature field. We represent this dependence with the two-dimensional heat-source layout temperature-field prediction (HSL-TFP) problem in Fig.~\ref{fig:problem_abstraction}. This abstraction originates from a general surrogate-modeling benchmark that represents distributed components as planar heat sources and predicts the resulting steady-state temperature field \citep{chen2021heatlayout}. Our previous work adapted this benchmark to flat-panel spacecraft by introducing heterogeneous component geometries, variable heat-rejection interfaces, and a broader thermal-parameter space \citep{wang2026dssl}. The present setting further increases compositional complexity through component counts of $K=1$--$25$, variable boundary-segment organizations, and the Temp-BC and RadFlux-BC scenarios, while retaining the principal physics of equipment heat dissipation, in-plane conduction, and boundary heat transfer.

\subsection{Task definition and input representation}

We consider steady thermal-field prediction on a two-dimensional modular spacecraft panel. Each sample is defined by the coordinate distribution $G$ on a computational domain $\Omega$, an internal heat-source component set $\mathcal{C}=\{c_i\}_{i=1}^{K}$, and a boundary-condition set $\mathcal{B}=\{b_j\}_{j=1}^{N_b}$. The model predicts a steady-state temperature field $T = \mathcal{H}(G,\mathcal{C},\mathcal{B}), \quad T\in\mathbb{R}^{H\times W\times 1}.$
Both $K$ and $N_b$ vary across samples, so evaluation focuses on extrapolation under increased component count.

\subsection{Boundary physics}

For the same modular spacecraft panel, this study considers two boundary condition scenarios: Prescribed-Temperature BC (Temp-BC) and Radiative-Flux BC (RadFlux-BC). The two scenarios use the same component-layout sampling strategy, but differ in the boundary physics used to generate the steady-state temperature fields. The governing steady heat-conduction equation and the two boundary conditions are summarized as:
\begin{equation}
\begin{aligned}
    -\nabla\cdot(k\nabla T)&=q(\mathbf{x};\mathcal{C}), && \mathbf{x}\in\Omega,\\
    T(\mathbf{x})&=T_b(\mathbf{x}), && \mathbf{x}\in\Gamma_D \quad \text{(Temp-BC)},\\
    -k\nabla T\cdot \mathbf{n}&=\epsilon\sigma\left(T^4-T_{\infty}^4\right),
    && \mathbf{x}\in\Gamma_R \quad \text{(RadFlux-BC)} .
\end{aligned}
\end{equation}
Here $q(\mathbf{x};\mathcal{C})$ is induced by the internal heat-source set. Temp-BC provides a direct boundary-temperature constraint, whereas RadFlux-BC introduces stronger source--boundary nonlinear coupling through the fourth-power radiation term.

\subsection{Data construction and evaluation protocol}

\begin{figure}[H]
\centering
\includegraphics[width=0.82\linewidth]{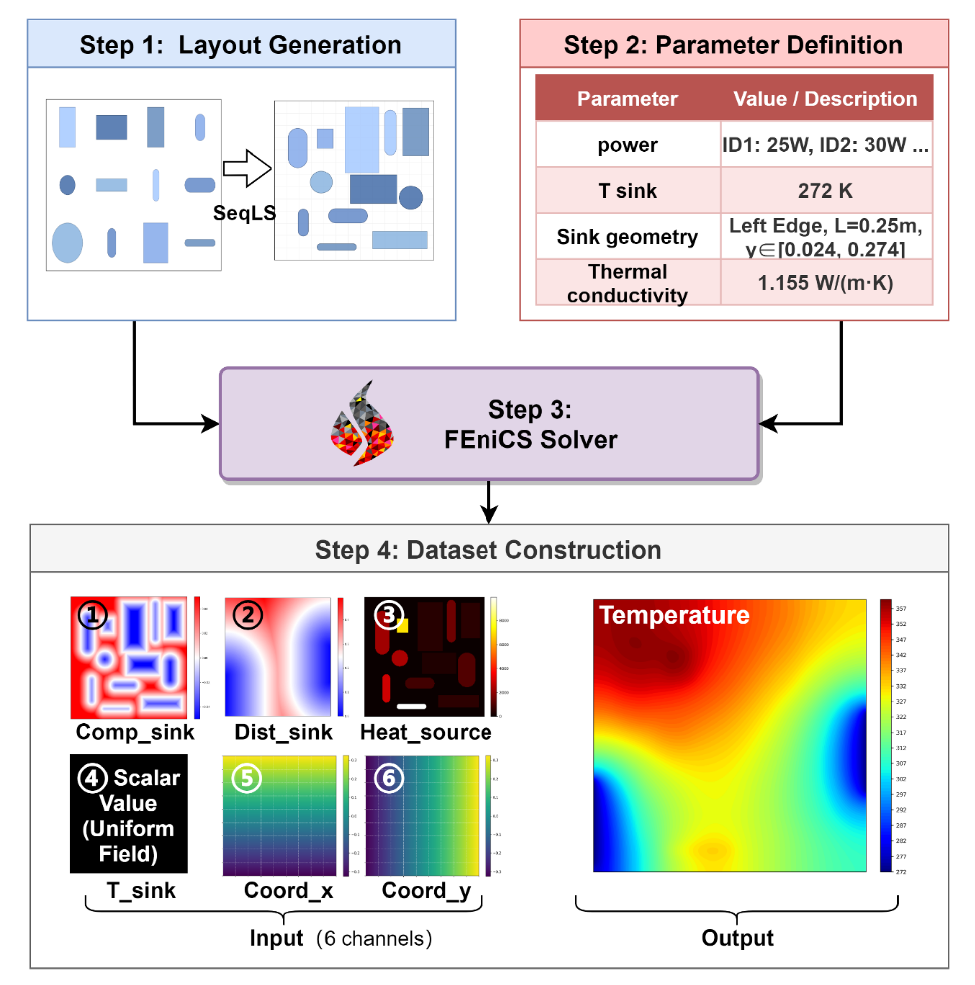}
\caption{Automated data-generation pipeline inherited from our previous work \citep{wang2026dssl}.}
\label{fig:data_generation_pipeline}
\end{figure}

Heat-source components and boundary segments are first provided in the raw data as spatial supports, positions, and physical parameters. Before they enter TFCN, both are converted into the unified factor representation described in Section~\ref{subsec:factor_representation}, namely SDF channels and normalized physical scalars. In implementation, $d_i$ and $\mathbf{v}_i$ are derived from raw heat-source components or boundary segments, stored separately as SDF channels and physical scalars, and reused through the same input interface in shared-parameter composition (TFCN-SP) and type-parameterized composition (TFCN-TP).
The data-generation procedure inherits the automated pipeline developed in our previous work, as illustrated in Fig.~\ref{fig:data_generation_pipeline}. Sequential Layout Sampling (SeqLS) first generates non-overlapping arrangements of square, circular, and capsule-shaped components. Physical parameters are then assigned within the previously established ranges, including domain side lengths of $0.5$--$2.0$~m, component powers of $5$--$50$~W, heat-sink temperatures of $260$--$290$~K, and a substrate thermal conductivity of $1.155$~W/(m$\cdot$K) \citep{wang2026dssl}. Given each layout--parameter configuration, the FEniCSx solver \citep{baratta2023dolfinx} computes the steady-state temperature field. The resulting sample pairs contain six input channels---component and heat-sink SDFs, the heat-source field, the heat-sink temperature field, and two coordinate grids---together with the temperature field as the output. The present study retains this four-stage generation procedure while reorganizing the samples according to component count and extending the boundary setting to the Temp-BC and RadFlux-BC scenarios defined in Section~2.2.

\begin{figure}[htbp]
\centering
\includegraphics[width=0.90\linewidth]{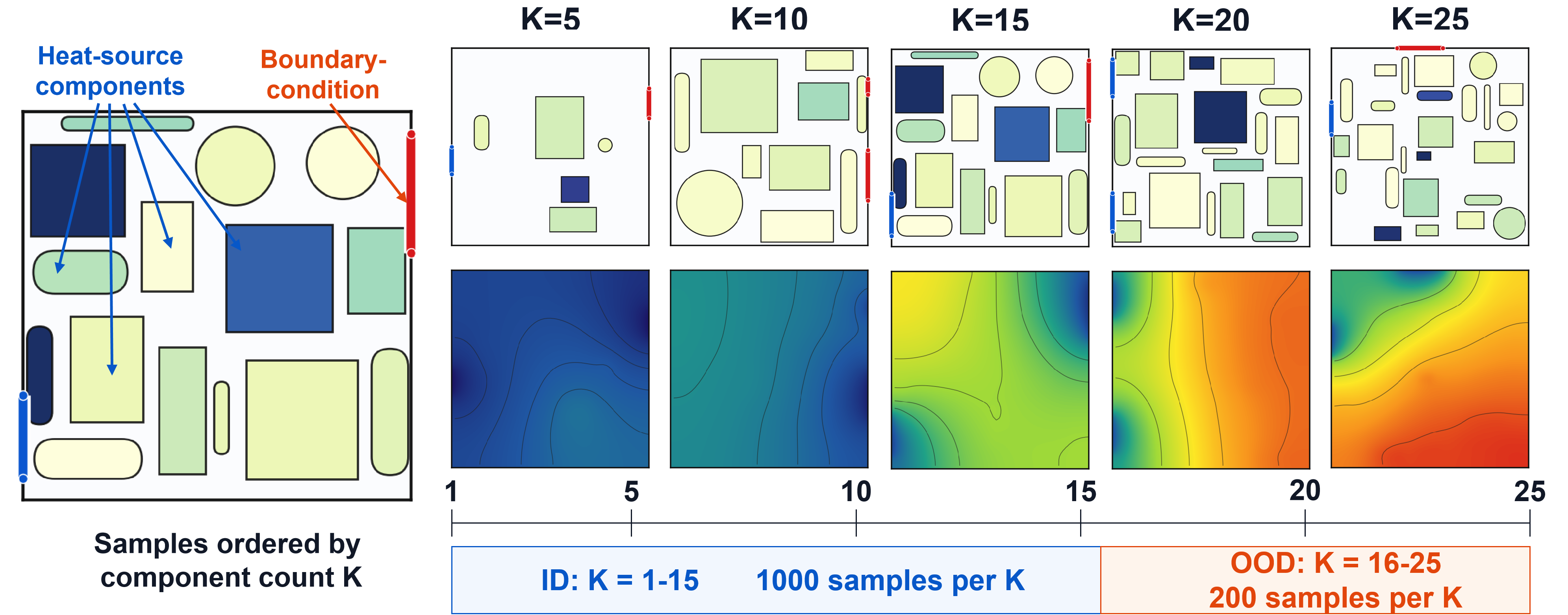}
\caption{Data construction and evaluation splits.}
\label{fig:dataset_protocol_full}
\end{figure}

Fig.~\ref{fig:dataset_protocol_full} summarizes data construction and evaluation splits. Samples are ordered by the number of internal heat-source components $K$. The component-count in-distribution split, denoted as $K$-\id, contains samples with $K=1\sim15$, whereas the component-count out-of-distribution split, denoted as $K$-\ood, contains samples with $K=16\sim25$. Each sample contains internal heat-source components, boundary-condition segments, and the corresponding steady-state temperature field.

The default boundary setting contains two boundary-condition segments, denoted as BC2.

\section{Method}
\label{sec:method}

Spacecraft thermal configuration design has an explicit compositional structure. Adding, replacing, or relocating onboard equipment changes the number, spatial arrangement, and power parameters of heat-source components, while modifications to the thermal-control scheme alter boundary segments and their physical parameters. Although each configuration defines a fully coupled thermal problem, different configurations are assembled from recurring engineering objects. We therefore formulate the task as learning a mapping from a variable set of geometry-bearing source and boundary factors to the complete temperature field. TFCN constructs the system-level latent representation through shared recursive composition. The following subsections present the overall framework, factor representation and bucket encoding, tree composition and field decoding, and the training and inference strategies.

\subsection{Overall framework: shared recursive composition of engineering factors}

Appendix Table~\ref{tab:principal_notation} summarizes the principal notation used in TFCN.

TFCN constructs the system-level representation through shared recursive composition of local engineering factors. Heat-source components and boundary segments are represented by their spatial supports and physical scalars, organized into balanced local groups, and encoded as latent tokens. A parameter-shared pairwise composer recursively merges these partial-configuration tokens, producing the representation used for full-field decoding.

Factorization is applied to the input objects, while their coupled thermal responses are learned jointly by the nonlinear composer and field decoder. Intermediate tree nodes encode latent partial configurations. The balanced binary tree specifies a regular, parameter-shared computation schedule and is not intended to represent component adjacency or physical heat-flow topology.

\begin{figure}[htbp]
\centering
\includegraphics[width=0.95\linewidth]{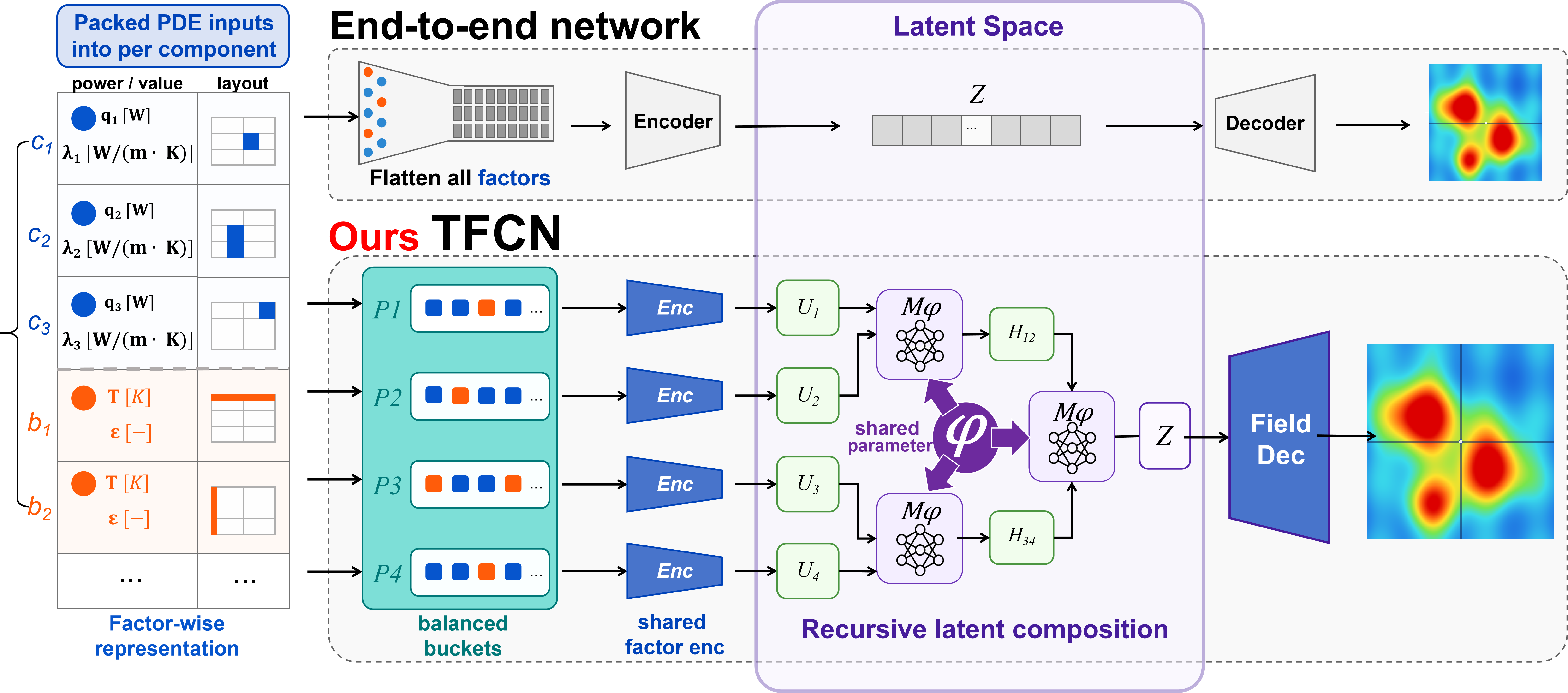}
\caption{Overall TFCN framework.}
\label{fig:pipeline}
\end{figure}

As shown in Fig.~\ref{fig:pipeline}, TFCN follows a factorize--bucketize--encode--compose--decode process. Let the heat-source components and boundary segments jointly form the factor set $\mathcal{F}=\{f_i\}_{i=1}^{n_f}$. One forward pass is written as
\begin{equation}
    \hat{Y}_{\mathrm{norm}}
    =
    \left[
    D_{\psi}(\,\cdot\,,G)
    \circ
    \mathcal{T}_{\phi}^{(D)}
    \circ
    E_{\theta}^{[Q]}
    \circ
    \Pi_Q
    \right](\mathcal{F}),
    \qquad Q=2^D.
\end{equation}
Here, $\Pi_Q$, $E_{\theta}^{[Q]}$, $\mathcal{T}_{\phi}^{(D)}$, and $D_{\psi}$ denote factor bucketing, shared bucket encoding, recursive composition, and field decoding, respectively. Their implementations are described in the following subsections.

\subsection{Factor representation and bucket encoding}
\label{subsec:factor_representation}

TFCN first organizes different numbers of heat-source components and boundary segments into a unified encodable factor interface. For the $i$-th heat-source component or boundary segment, its spatial support is denoted as $s_i\subset\Omega$, and its signed distance field (SDF) and factor representation are jointly defined as:
\begin{equation}
\begin{aligned}
    d_i(\mathbf{x})
    &=\mathrm{SDF}(s_i)(\mathbf{x}),\quad \mathbf{x}\in\Omega,\\
    f_i
    &=(d_i,\mathbf{v}_i),\quad
    \mathbf{v}_i\in\mathbb{R}^{p_i}.
\end{aligned}
\end{equation}
This study adopts the sign convention that the SDF is negative inside the spatial support and positive outside it. The physical value $\mathbf{v}_i$ is normalized according to the corresponding physical quantity, such as heat-source power, boundary temperature, heat-transfer intensity, or radiation-related parameters. In the experiments of this study, each source factor and BC factor uses one normalized scalar as its value feature, namely $p_i=1$. This representation stores spatial support and physical strength separately: the SDF provides a continuous distance to the support boundary and avoids the coarsening caused by binary masks near boundaries, while $\mathbf{v}_i$ provides source or boundary-condition strength and avoids direct mixing of raw values with different dimensions. No additional factor-type channel is used in the present implementation. The two variants differ in how factor types are handled during bucketing and encoding.

After factor-level SDFs and value features are obtained, TFCN performs bucketing. Given a factor set $\mathcal{F}=\{f_i\}_{i=1}^{n_f}$ that participates in the current forward pass and a composition depth $D$, the model creates $Q=2^D$ buckets and uses the bucketing operator $\Pi_Q$ to form the pre-encoding bucket-feature set:
\begin{equation}
    Q=2^D,\qquad
    \mathcal{X}=\{X_j\}_{j=1}^{Q}=\Pi_Q(\mathcal{F}).
\end{equation}
The bucketing operator $\Pi_Q$ randomly permutes factors without spatial or physical sorting. At each forward pass, it assigns the $r$-th permuted factor to bucket $j=1+((r-1)\bmod Q)$, so bucket sizes differ by at most one. A new permutation is sampled during both training and inference. For reproducible evaluation, the inference random state and number of sampled paths are fixed. Random rebucketing changes only factor grouping and composition order; all paths share the same trained parameters.
Empty factor slots are handled using fixed padding values and binary validity masks. The complete tensor construction is provided in \ref{app:factor_padding}. In the present implementation, the input width of 20 denotes the source-factor capacity of each bucket, rather than the maximum component count of the complete configuration. Since the model uses $Q=2^D$ buckets, the total source-factor capacity is $20Q$; under the default setting $Q=4$, a configuration with $K=25$ contains at most $\lceil25/4\rceil=7$ components in each bucket and is therefore represented without truncation.

In TFCN, heat-generating components and boundary segments are both organized through the same factor interface: each factor is described by a layout SDF and normalized value features. Source factors represent energy injection inside the domain, whereas BC factors represent boundary constraints, boundary heat exchange, or radiation-related conditions.

Let $\mathcal{F}_c$ and $\mathcal{F}_b$ denote the source-factor set derived from $\mathcal{C}$ and the BC-factor set derived from $\mathcal{B}$, respectively. \sharedparam\ places source and BC factors in mixed buckets and processes them with a shared encoder. \typedparam\ applies type-separated bucketing and independent source/BC encoders. In both variants, the resulting token sets enter the same tree composer and decoder. The distinction between the two variants is confined to bucketing and encoder parameterization.

The two paths can be written compactly as:
\begin{equation}
\label{eq:boundary_organization}
\begin{aligned}
\sharedparam:\quad
&\mathcal{U}_{\mathrm{SP}}
=\bigl(E_{\theta_s}^{[Q]}\circ\Pi_Q\bigr)(\mathcal{F}_c\cup\mathcal{F}_b),\\
\mathcal{X}_{\mathrm{SP}}
&=\{X^{\mathrm{SP}}_j\}_{j=1}^{Q}
=\Pi_Q(\mathcal{F}_c\cup\mathcal{F}_b),\\
\mathcal{U}_{\mathrm{SP}}
&=\{E_{\theta_s}(X^{\mathrm{SP}}_j)\}_{j=1}^{Q},\\
\typedparam:\quad
&\mathcal{U}_{\mathrm{TP}}
=\bigl(E_{\theta_c}^{[Q_c]}\circ\Pi_{Q_c}^{c}\bigr)(\mathcal{F}_c)
\cup
\bigl(E_{\theta_b}^{[Q_b]}\circ\Pi_{Q_b}^{b}\bigr)(\mathcal{F}_b),\\
\mathcal{X}_{c}
&=\{X^{c}_j\}_{j=1}^{Q_c}
\!=\Pi_{Q_c}^{c}(\mathcal{F}_c),\\
\mathcal{X}_{b}
&=\{X^{b}_j\}_{j=1}^{Q_b}
=\Pi_{Q_b}^{b}(\mathcal{F}_b),\\
\mathcal{U}_{\mathrm{TP}}
&=\{E_{\theta_c}(X^{c}_j)\}_{j=1}^{Q_c}
\cup
\{E_{\theta_b}(X^{b}_j)\}_{j=1}^{Q_b}.
\end{aligned}
\end{equation}
Here $\Pi_Q$ denotes the mixed bucketing operator. $\Pi_{Q_c}^{c}$ and $\Pi_{Q_b}^{b}$ denote the type-separated source and BC bucketing operators, respectively, where $Q_c$ and $Q_b$ allocate the total bucket budget $Q$ between source and BC factors. $\mathcal{X}_{\mathrm{SP}}$, $\mathcal{X}_{c}$, and $\mathcal{X}_{b}$ are all pre-encoding bucket-feature sets. $E_{\theta_s}^{[Q]}$, $E_{\theta_c}^{[Q_c]}$, and $E_{\theta_b}^{[Q_b]}$ denote applying the corresponding encoder to several bucket features: \sharedparam\ uses shared parameters $\theta_s$ for mixed bucket features, whereas \typedparam\ uses independent parameters $\theta_c$ and $\theta_b$ for source and BC bucket features. $\mathcal{U}_{\mathrm{SP}}$ and $\mathcal{U}_{\mathrm{TP}}$ are token sets before the composition stage.

\subsection{Tree composer and field decoder}

Each encoded token represents a local factor group, or equivalently a partial configuration in the latent space. TFCN uses a shared pairwise composer to map two partial-configuration representations to the representation of their joint configuration:
\begin{equation}
    M_{\phi}(U,V)=U+V+R_{\phi}([U,V]),
\end{equation}
where $U+V$ provides a low-order additive path and $R_{\phi}$ is a residual convolutional network that learns corrections associated with source--source interactions, source--boundary coupling, geometric proximity, and nonlinear boundary physics.

The tree composer merges tokens layer by layer along a binary tree. Layer $0$ is $\{U_j\}_{j=1}^{Q}$, where $Q=2^D$, and layer $\ell+1$ is obtained by merging adjacent tokens in layer $\ell$:
\begin{equation}
    U^{(\ell+1)}_q
    =
    M_{\phi}\!\left(U^{(\ell)}_{2q-1},U^{(\ell)}_{2q}\right),
    \quad
    \ell=0,\ldots,D-1,\quad
    q=1,\ldots,2^{D-\ell-1}.
\end{equation}
All tree layers share the same $M_{\phi}$, so the number of parameters does not grow with component count or bucket count. The final composed representation is denoted as $Z$.

In the spatial implementation, the local composer operates on disjoint $4\times4$ spatial windows of encoded latent maps $U,V\in\mathbb{R}^{H\times W\times d_z}$. Because $H$ and $W$ are divisible by 4, corresponding window pairs are composed independently by the shared $M_\phi$ and stitched back at their original positions without spatial padding. This partition acts only on spatial dimensions and is unrelated to factor-token indices or ordering. The w/o Locality ablation instead merges each pair over the complete latent maps.

The decoder reconstructs the complete normalized temperature field $\hat{Y}_{\mathrm{norm}}$ from the composed representation $Z$. The coordinate distribution $G$ supplies positional information to the decoder for field reconstruction.

\subsection{Training and inference}

The training strategy maintains a fixed tensor interface while varying composition granularity. At each training step, the depth is sampled as $D\sim\mathrm{Uniform}\{0,1,\ldots,D_{\max}\}$; the default is $D_{\max}=2$, with $D_{\max}=0,1,3$ examined in the structural analysis.

During evaluation, six inference paths are generated for each input, including two independently constructed bucket assignments at each composition depth ($D\in\{1,2,3\}$). Under the prescribed evaluation random states, these paths share the same network parameters and can be evaluated in parallel. DBSCAN \citep{ester1996dbscan} is applied to the resulting latent representations $Z^{(p)}$; candidate representations identified as noise by DBSCAN are excluded, and the remaining representations indexed by $\mathcal{I}$ are combined by element-wise arithmetic averaging before decoding. This procedure is used as a fixed evaluation-time aggregation rule rather than an additional learned module.
\begin{equation}
    \bar{Z}=\frac{1}{|\mathcal{I}|}\sum_{p\in\mathcal{I}}Z^{(p)}.
    \label{eq:multipath_mean}
\end{equation}
The shared decoder processes the fused representation $\bar{Z}$ once to obtain the final field.

\begin{figure}[htbp]
\centering
\includegraphics[width=0.95\linewidth]{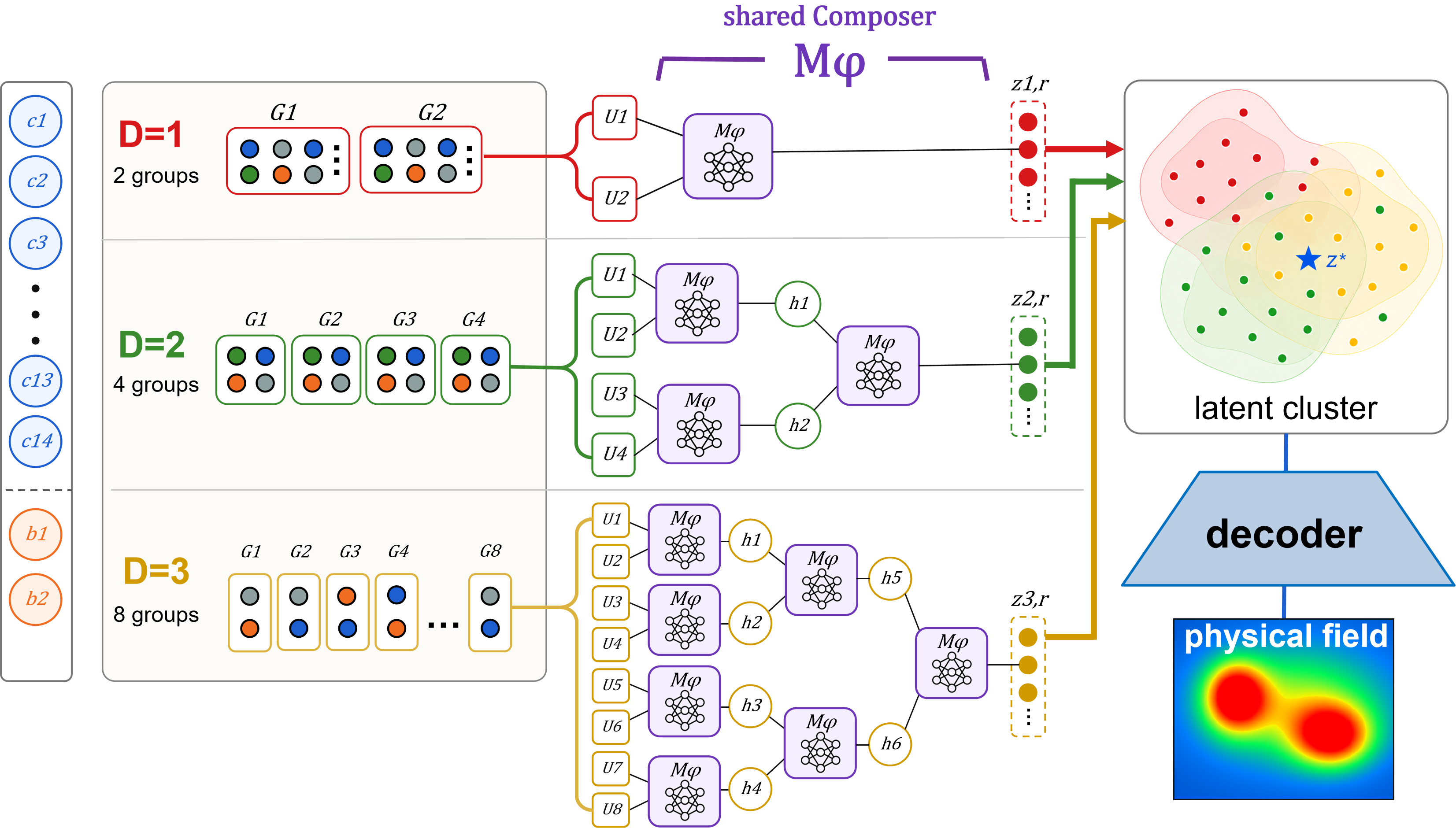}
\caption{Multi-path latent-space composition.}
\label{fig:tree_composition}
\end{figure}

The training loss is the mean squared error on the normalized temperature field. For a physical temperature field $T$, its normalized label is denoted as $Y_{\mathrm{norm}}=\mathcal{N}(T)$, and the network prediction is $\hat{Y}_{\mathrm{norm}}$:
\begin{equation}
\begin{aligned}
    \Theta&=\{\theta_s,\theta_c,\theta_b,\phi,\psi,\ldots\},\\
    \mathcal{L}(\Theta)
    &=
    \frac{1}{|\mathcal{D}|HW}
    \sum_{\xi\in\mathcal{D}}
    \sum_{\mathbf{x}\in\Omega_h}
    \left(\hat{Y}_{\mathrm{norm}}^{\xi}(\mathbf{x})
    -Y_{\mathrm{norm}}^{\xi}(\mathbf{x})\right)^2 .
\end{aligned}
\end{equation}
Here $\xi=(G,\mathcal{C},\mathcal{B},Y_{\mathrm{norm}})$ denotes one training sample. During evaluation, predictions are inverse-normalized to physical units, and RMSE and MAE are computed over the corresponding evaluation set.

During evaluation, the inference random state and number of sampled paths are controlled for reproducibility.

Together, random-depth training and controlled multi-path inference expose the shared encoder and composer to different factor groupings while preserving a fixed prediction interface.

\section{Experiments and analysis}
\label{sec:experiments}

\subsection{Setup}

The experiments assess TFCN under $K$-\ood\ extrapolation, changes in training coverage and composition depth, structural ablations, and diagnostics of intermediate representations. These analyses examine prediction performance at high component counts and its dependence on training coverage and composer structure.

The comparison includes a convolutional U-Net \citep{ronneberger2015unet}, the branch--trunk operator architecture DeepONet \citep{lu2021deeponet}, and the geometry-informed neural operator GINO \citep{li2023gino} as representative global and operator-learning baselines.

\subsection{Main results}

\begin{table}[htbp]
\caption{Interpolation and high-component-count extrapolation performance. Values are reported as mean $\pm$ standard deviation over five independent runs.}
\label{tab:main_results}
\vspace{0.05em}
\centering
\footnotesize
\setlength{\tabcolsep}{5.5pt}
\begin{tabular}{llcccc}
\toprule
\multirow{2}{*}{Dataset} & \multirow{2}{*}{Method}
& \multicolumn{2}{c}{$K$-ID}
& \multicolumn{2}{c}{$K$-OOD} \\
\cmidrule(lr){3-4}\cmidrule(lr){5-6}
& & RMSE & MAE & RMSE & MAE \\
\midrule
\multirow{5}{*}{Temp-BC}
& UNet & $9.99 \pm 0.27$ & $6.83 \pm 0.10$ & $46.91 \pm 1.56$ & $33.52 \pm 1.22$ \\
& DeepONet & $12.94 \pm 0.24$ & $9.18 \pm 0.12$ & $38.95 \pm 1.19$ & $29.13 \pm 0.68$ \\
& GINO & $3.09 \pm 0.15$ & $2.06 \pm 0.04$ & $12.23 \pm 0.55$ & $5.42 \pm 0.32$ \\
& TFCN-SP & $4.88 \pm 0.76$ & $2.97 \pm 0.42$ & $15.66 \pm 2.00$ & $9.12 \pm 2.37$ \\
& TFCN-TP & $\boldsymbol{1.45 \pm 0.13}$ & $\boldsymbol{0.95 \pm 0.07}$ & $\boldsymbol{4.21 \pm 0.15}$ & $\boldsymbol{2.56 \pm 0.09}$ \\
\midrule
\multirow{5}{*}{RadFlux-BC}
& UNet & $18.77 \pm 0.44$ & $14.23 \pm 0.27$ & $72.16 \pm 2.75$ & $53.09 \pm 2.26$ \\
& DeepONet & $23.08 \pm 0.53$ & $18.23 \pm 0.36$ & $63.12 \pm 1.96$ & $46.77 \pm 1.78$ \\
& GINO & $9.11 \pm 0.66$ & $6.21 \pm 0.51$ & $28.05 \pm 0.68$ & $22.58 \pm 0.55$ \\
& TFCN-SP & $10.53 \pm 1.43$ & $7.32 \pm 1.06$ & $23.06 \pm 2.65$ & $15.45 \pm 2.49$ \\
& TFCN-TP & $\boldsymbol{8.78 \pm 1.76}$ & $\boldsymbol{5.86 \pm 1.12}$ & $\boldsymbol{18.62 \pm 7.50}$ & $\boldsymbol{15.11 \pm 6.60}$ \\
\bottomrule
\end{tabular}
\end{table}

Table~\ref{tab:main_results} reports the main interpolation and high-component-count extrapolation results under Temp-BC and RadFlux-BC. All entries are reported as the mean $\pm$ standard deviation over five independent runs with different random seeds. Both TFCN variants outperform UNet and DeepONet in mean $K$-\ood\ error, and the best TFCN result in each setting also outperforms GINO. In Temp-BC, TFCN-TP achieves the lowest mean $K$-\ood\ \rmse\ of $4.21$ K, compared with $12.23$ K for GINO, $46.91$ K for UNet, and $38.95$ K for DeepONet. In RadFlux-BC, TFCN-TP achieves the lowest mean $K$-\ood\ \rmse\ of $18.62$ K, compared with $28.05$ K for GINO, $72.16$ K for UNet, and $63.12$ K for DeepONet. These results show that factor composition improves mean high-component-count extrapolation accuracy over the conventional global surrogates and the geometry-informed operator baseline.

To complement field-wise \rmse\ and \mae, Appendix Tables~\ref{tab:thermal_indicator_definitions} and~\ref{tab:thermal_indicator_results} define and report four indicators of peak-temperature accuracy and tail risk, hot-spot localization, and boundary-band fidelity; TFCN-TP achieves the lowest ID/OOD errors for all four under both boundary settings.

\begin{figure}[H]
\centering
\begin{minipage}{0.99\linewidth}
\centering

\begin{minipage}{0.49\linewidth}
\centering
\includegraphics[width=\linewidth]{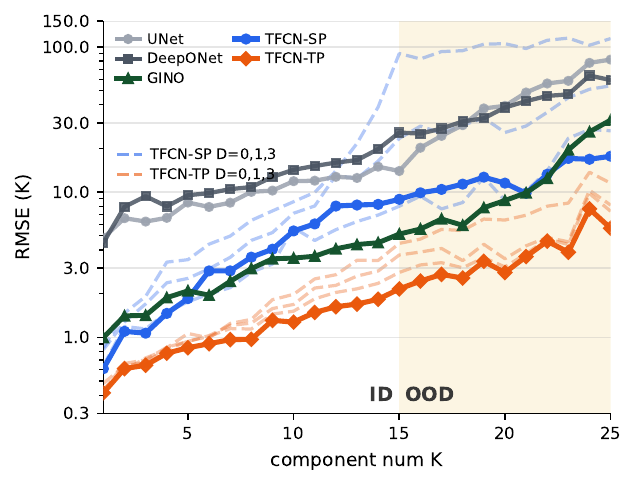}
\par\vspace{0.1em}{\small (a) Temp-BC}
\end{minipage}
\hfill
\begin{minipage}{0.49\linewidth}
\centering
\includegraphics[width=\linewidth]{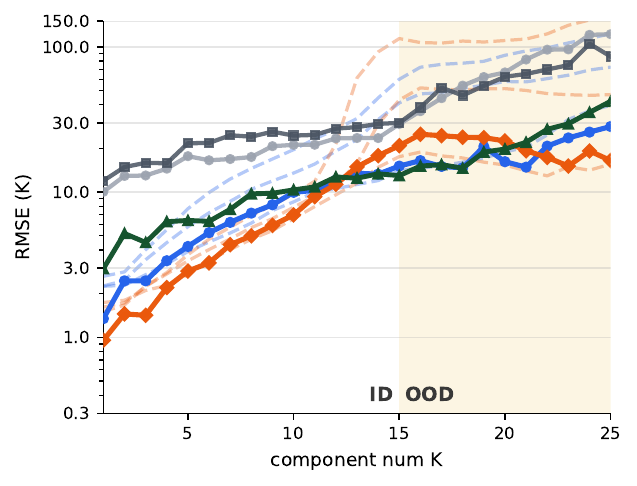}
\par\vspace{0.1em}{\small (b) RadFlux-BC}
\end{minipage}

\end{minipage}
\caption{Per-$K$ extrapolation error.}
\label{fig:k_extrapolation}
\end{figure}

The two TFCN variants differ in how source and boundary factors are parameterized. TFCN-TP achieves the lowest mean RMSE and MAE in both boundary-condition settings. Its advantage over TFCN-SP is larger in Temp-BC, while the RadFlux-BC results show greater run-to-run variability, as indicated by the larger standard deviations. These results indicate that factor parameterization affects both mean accuracy and stability.

After decomposing errors by component count $K$, the model differences in the extrapolation interval become clearer. Fig.~\ref{fig:k_extrapolation} uses a logarithmic vertical axis, so vertical separation corresponds to multiplicative error differences rather than simple linear gaps. After entering the $K=16\sim25$ extrapolation interval, the UNet and DeepONet curves rise rapidly in both datasets; in RadFlux-BC, their per-$K$ \rmse\ can reach the order of $150$ K. In contrast, the TFCN curves grow more gradually, and the main TFCN variants in RadFlux-BC stay below about $30$ K at most high-$K$ positions. The performance advantage persists across most component-count groups in the OOD interval.

Appendix Fig.~\ref{fig:qualitative_multi} maps the scalar error advantage to temperature-field structure. In high-component-count RadFlux-BC samples, TFCN preserves the main hot-spot locations, boundary-induced gradients, and local temperature-rise structures. Errors are mainly concentrated near boundary high-gradient regions and local heat-source neighborhoods, rather than forming broad field-wide shifts. Additional RadFlux-BC prediction samples are provided in Appendix Fig.~\ref{fig:qualitative_multi_appendix}. These field-level results complement the \rmse\ evaluation with qualitative evidence of preserved field structure.

\subsection{Training coverage, composition depth, and ablations}

\begin{figure}[!htbp]
\centering
\includegraphics[width=0.6\linewidth]{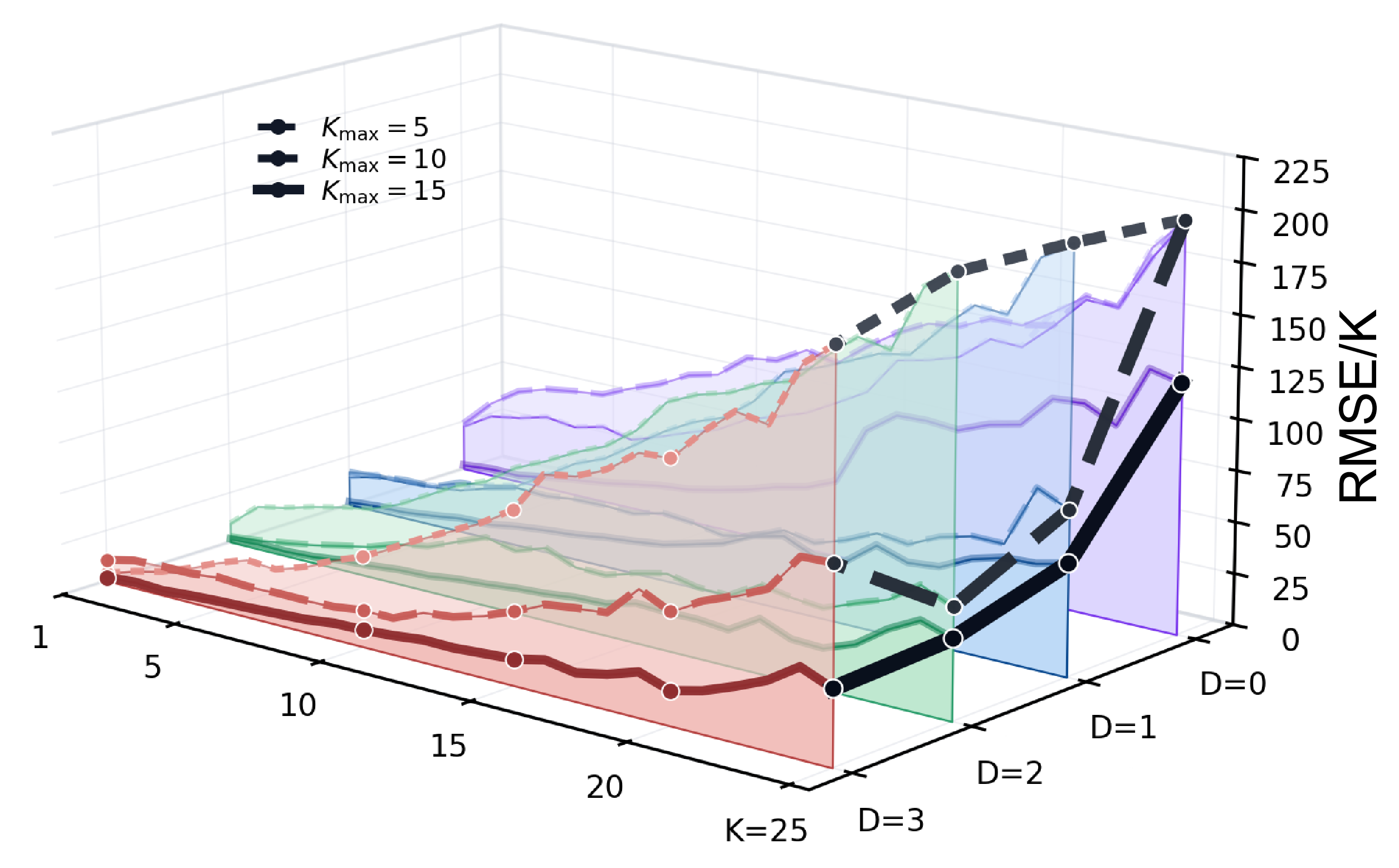}
\caption{Training coverage and composition depth.}
\label{fig:kmax_depth}
\end{figure} 

Fig.~\ref{fig:kmax_depth} evaluates the coupled effects of training coverage $K_{\max}$ and maximum composition depth $D_{\max}$ under the RadFlux-BC/TFCN-SP setting. The three curves in each depth slice correspond to $K_{\max}=5,10,15$, with the horizontal axis denoting test component count $K$ and the vertical axis denoting per-$K$ \rmse. This comparison separates the effect of training coverage from that of composition granularity.

Fig.~\ref{fig:kmax_depth} shows that increasing maximum composition depth alone cannot compensate for insufficient training coverage. When $K_{\max}=5$, the high-$K$ average errors under $D_{\max}=0,1,2,3$ remain at the order of hundreds of Kelvin, indicating that the training data do not provide enough local composition experience for reliable extrapolation. As coverage expands, composition depth becomes effective: for $K_{\max}=10$, $D_{\max}=2$ reduces the high-$K$ mean \rmse\ to about $40.06$ K, compared with $136.03$ K under $D_{\max}=0$; for $K_{\max}=15$, $D_{\max}=2$ and $D_{\max}=3$ further reduce the mean to about $24.36$ K and $22.33$ K, respectively. Because the main architecture, decoder interface, and parameter scale are held fixed, this comparison isolates the effects of training coverage and composition depth. The results indicate that TFCN uses composition experience from the training configurations when constructing unseen high-$K$ representations through the shared composer.

\begin{table}[htbp]
\caption{Structural ablation results.}
\label{tab:ablation}
\vspace{0.05em}
\centering
\small
\begin{tabular}{llrrrrr}
\toprule
\multirow{2}{*}{Dataset} & \multirow{2}{*}{Method}
& \multicolumn{2}{c}{$K$-ID}
& \multicolumn{2}{c}{$K$-OOD}
& \multirow{2}{*}{OOD/ID} \\
\cmidrule(lr){3-4}\cmidrule(lr){5-6}
& & RMSE & MAE & RMSE & MAE & \\
\midrule
\multirow{3}{*}{Temp-BC}
& TFCN-SP & 4.88 & 2.97 & \textbf{15.66} & \textbf{9.12} & 3.209 \\
& w/o Locality & \textbf{4.35} & \textbf{2.53} & 18.49 & 10.29 & 4.254 \\
& w/o Composition & 18.62 & 11.58 & 54.03 & 40.11 & 2.902 \\
\midrule
\multirow{3}{*}{RadFlux-BC}
& TFCN-SP & 10.53 & 7.32 & \textbf{23.06} & \textbf{15.45} & 2.190 \\
& w/o Locality & \textbf{10.00} & \textbf{7.12} & 25.29 & 15.70 & 2.529 \\
& w/o Composition & 42.98 & 22.00 & 110.57 & 85.99 & 2.573 \\
\bottomrule
\end{tabular}
\end{table}

Table~\ref{tab:ablation} compares TFCN-SP with two structural ablations. The w/o Composition variant replaces the learnable composer $M_{\phi}$ with fixed minimum aggregation of the input SDFs and direct fusion of scalar features. The w/o Locality variant retains the learnable composer and random balanced bucketing but applies each merge to the complete latent maps. In RadFlux-BC, removing local-block merging increases the $K$-\ood\ RMSE from $23.06$ K to $25.29$ K, whereas removing the learnable composer increases it to $110.57$ K. The corresponding Temp-BC $K$-\ood\ RMSE rises from $15.66$ K to $18.49$ K and $54.03$ K, respectively. These results indicate that the learnable composer provides the main OOD improvement, while local windowing offers an additional benefit.

\subsection{Intermediate-representation and field-level diagnostics}

PHATE is used to visualize the projected geometry of the \id\ and $K$-\ood\ intermediate representations \citep{moon2019phate}, while maximum mean discrepancy (MMD) quantifies their empirical distributional discrepancy \citep{gretton2012mmd}. These measures serve as descriptive diagnostics of intermediate-representation geometry and distribution.

At the distribution level, MMD compares two distributions through kernel mean embeddings in a reproducing-kernel Hilbert space:
\begin{equation}
    \mathrm{MMD}^2(P,Q)=
    \left\|\mathbb{E}_{z\sim P}\phi(z)-\mathbb{E}_{z'\sim Q}\phi(z')\right\|_{\mathcal{H}}^2 .
\end{equation}
Here $P$ and $Q$ denote the pre-decoder intermediate-representation distributions of the \id\ set and a $K$-\ood\ group, respectively, and $\phi(\cdot)$ is the kernel-induced feature map. The resulting MMD value measures the empirical discrepancy between the two intermediate-representation sets.

\begin{figure}[!htbp]
\centering
\begin{minipage}{0.8\linewidth}
\centering

\begin{minipage}{0.45\linewidth}
\centering
\includegraphics[width=\linewidth,trim=0 0 0 24bp,clip]{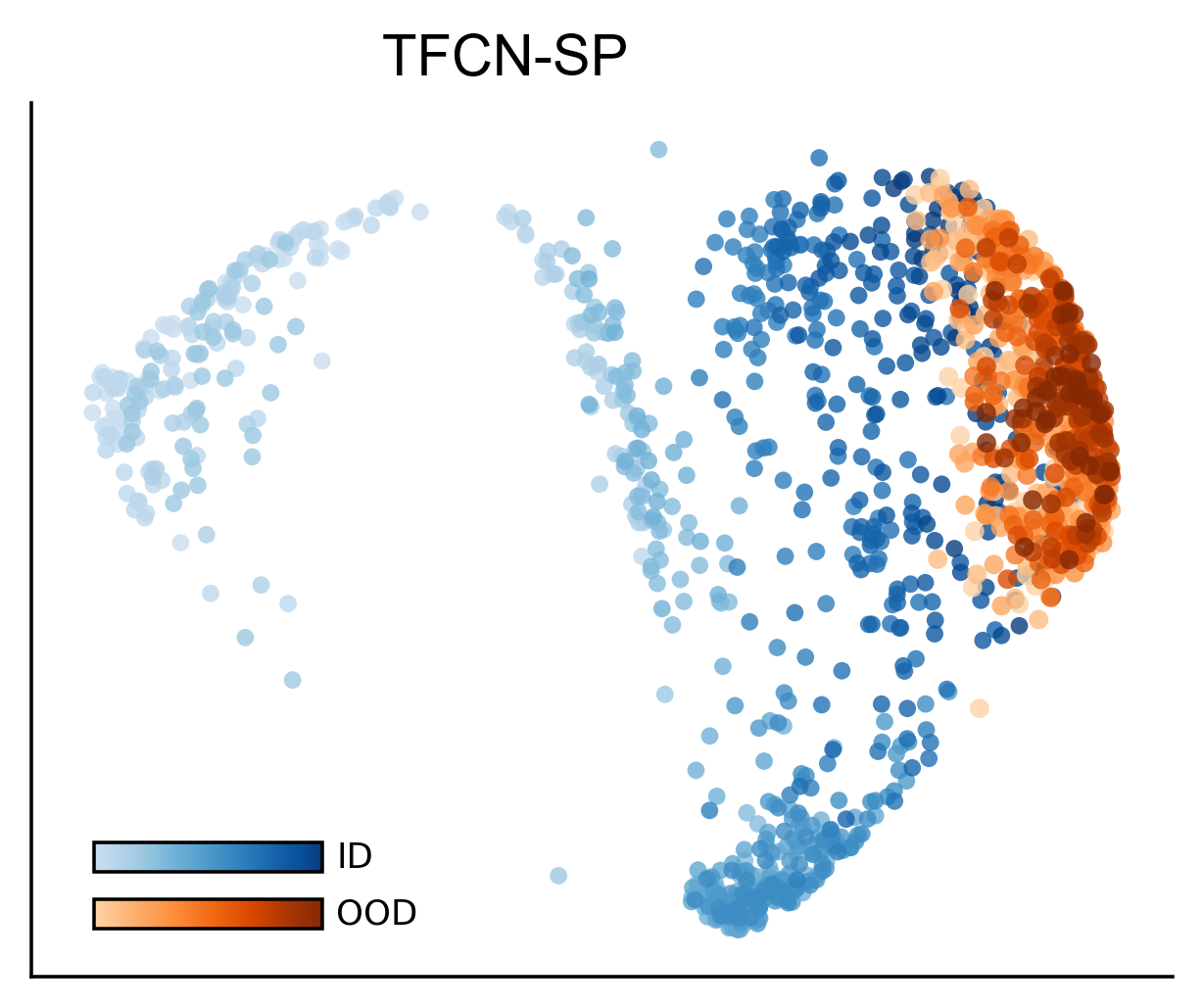}
\par\vspace{0.1em}{\small (a) TFCN-SP}
\end{minipage}
\hfill
\begin{minipage}{0.45\linewidth}
\centering
\includegraphics[width=\linewidth,trim=0 0 0 24bp,clip]{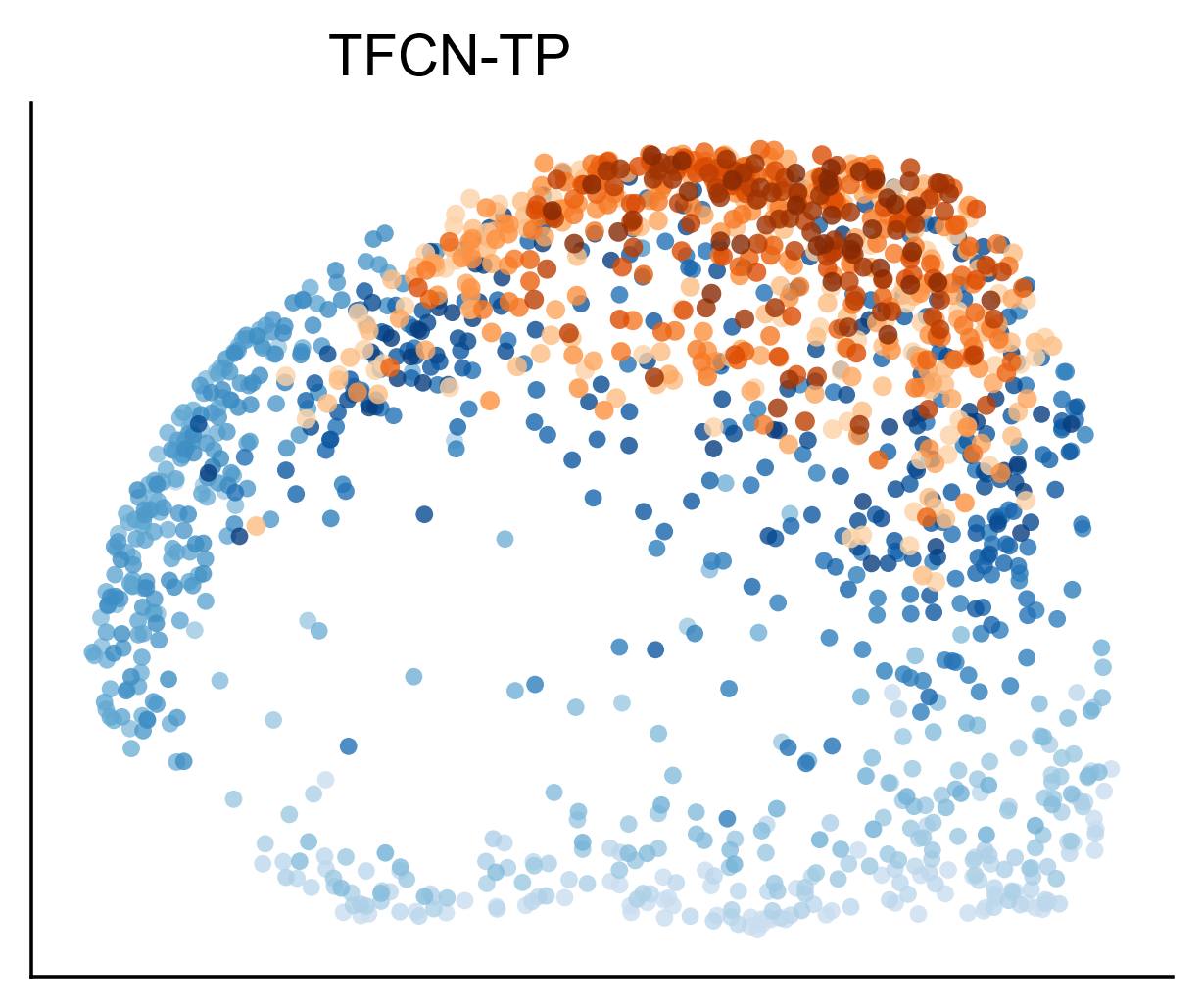}
\par\vspace{0.1em}{\small (b) TFCN-TP}
\end{minipage}

\vspace{0.5em}

\begin{minipage}{0.45\linewidth}
\centering
\includegraphics[width=\linewidth,trim=0 0 0 22bp,clip]{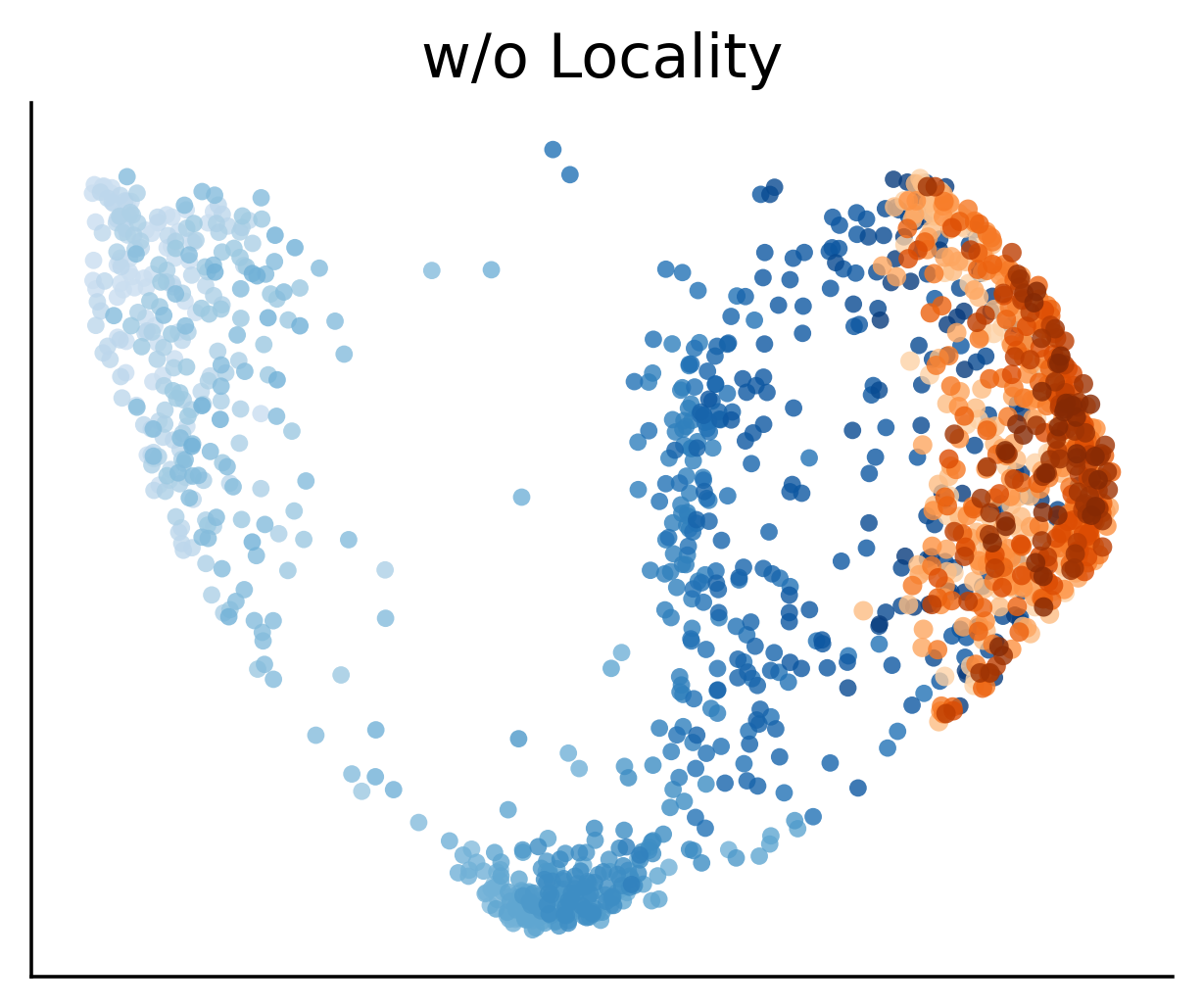}
\par\vspace{0.1em}{\small (c) w/o Locality}
\end{minipage}
\hfill
\begin{minipage}{0.45\linewidth}
\centering
\includegraphics[width=\linewidth,trim=0 0 0 22bp,clip]{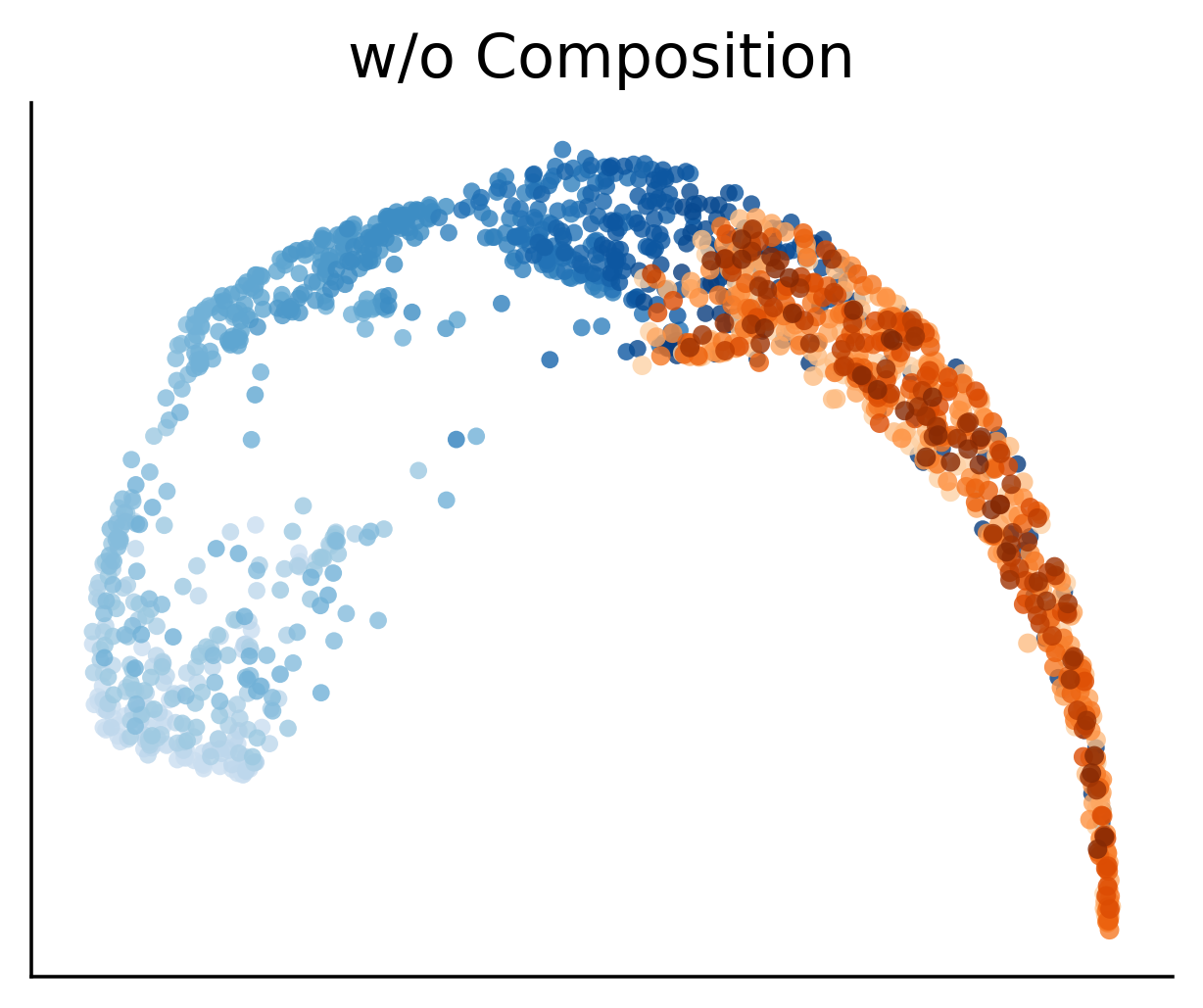}
\par\vspace{0.1em}{\small (d) w/o Composition}
\end{minipage}

\end{minipage}
\caption{PHATE diagnostic of intermediate-representation geometry.}
\label{fig:latent_phate_linear}
\end{figure}

Fig.~\ref{fig:latent_phate_linear} shows joint-fit PHATE projections of pre-decoder intermediate representations. Each point denotes a validation sample, with blue for \id\ and orange for $K$-\ood; color intensity indicates component count $K$. In the two-dimensional projection, OOD points for \sharedparam, \typedparam, and w/o Locality remain closer to the projected ID region, whereas w/o Composition shows a larger separation between the two groups. The color gradients also indicate that component count is reflected in the projected organization of the intermediate representations.

\begin{figure}[!htbp]
\centering
\includegraphics[width=0.8\linewidth]{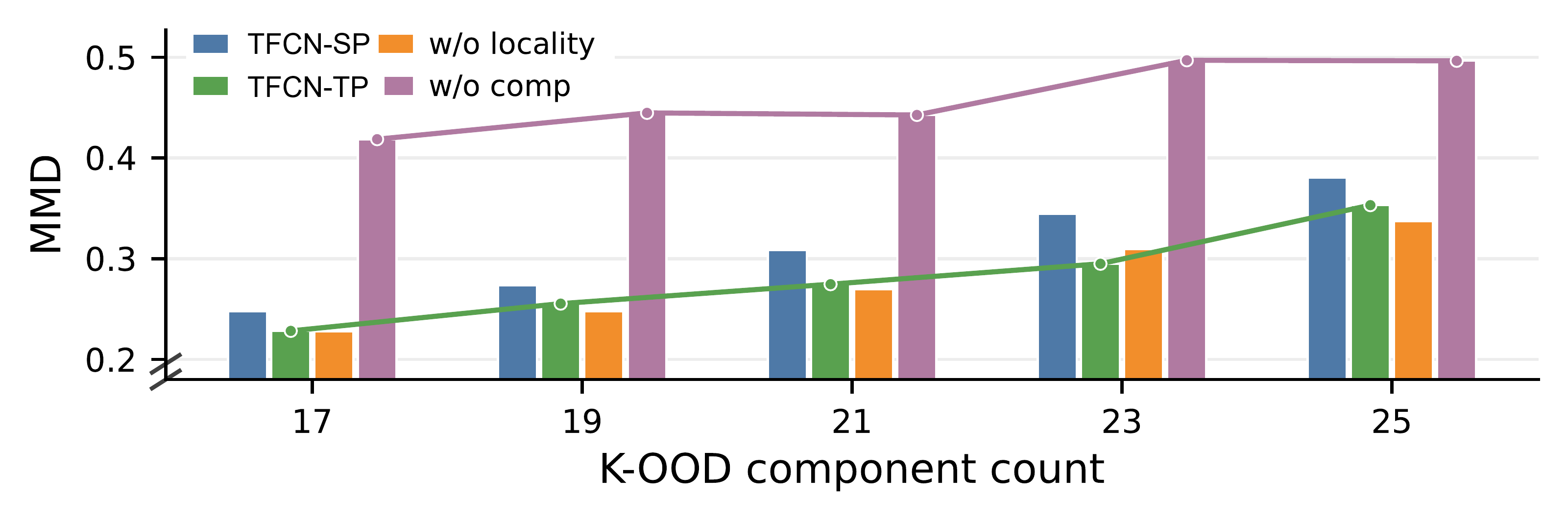}
\caption{MMD diagnostic of the ID/OOD latent-distribution discrepancy.}
\label{fig:mmd_k}
\end{figure}

At the distribution scale, Fig.~\ref{fig:mmd_k} reports the MMD between the ID representations and the representations at each odd high component count from $K=17$ to $K=25$ in the RadFlux-BC dataset. Across all high-$K$ groups, w/o Composition has the largest MMD. The remaining variants show closer MMD values, and the diagnostic does not establish a consistent ranking between locality and factor-parameterization choices. Together with the PHATE projections, these results show that removing the composer is associated with a larger empirical ID/OOD discrepancy in the intermediate representations.

\begin{figure}[!htbp]
\centering
\includegraphics[width=0.7\linewidth]{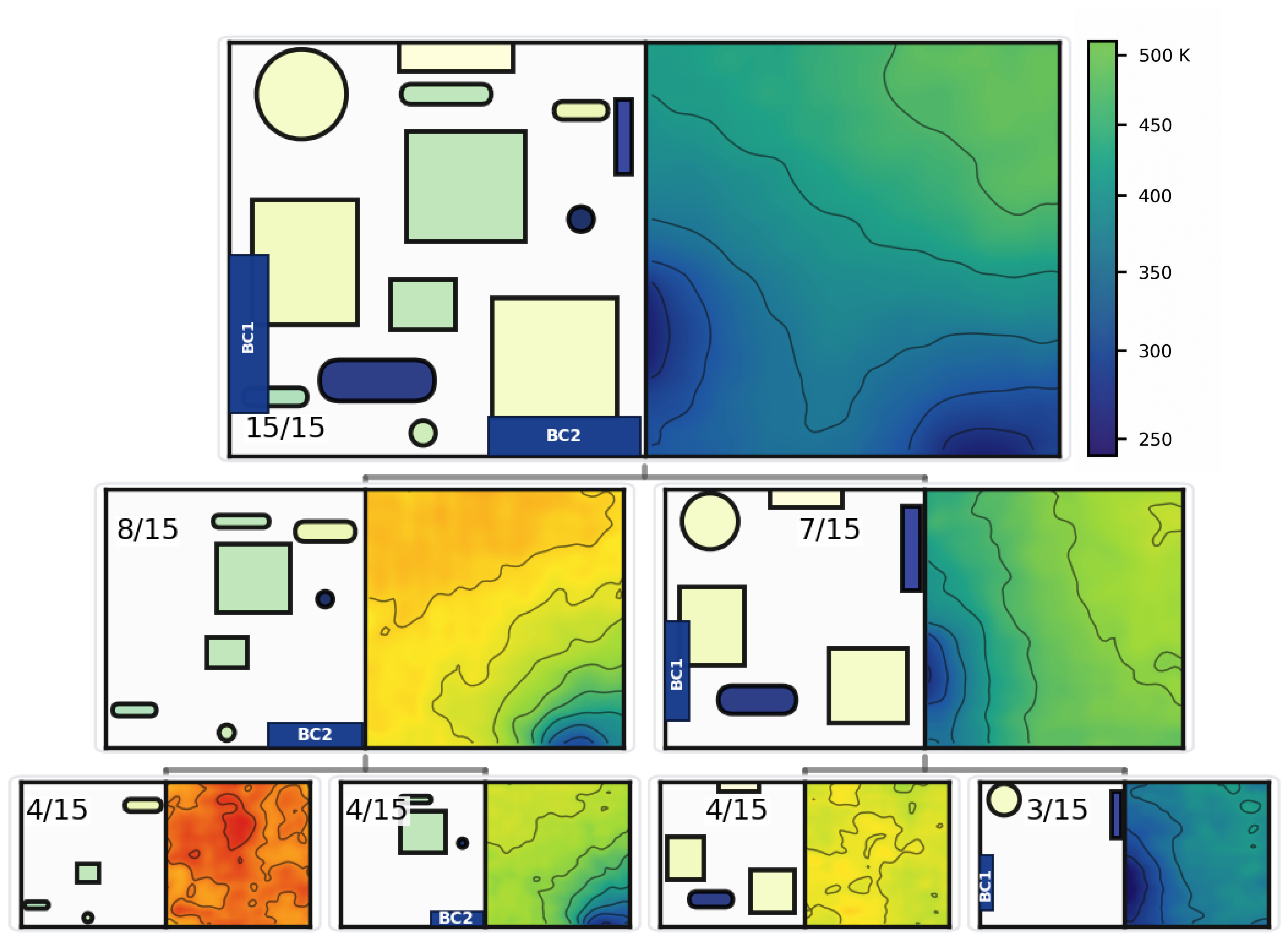}
\caption{Single-sample intermediate-representation diagnostic during factor composition. Intermediate panels are decoder projections; only the fully composed $15/15$ panel is the final full-system temperature prediction and uses the Kelvin scale.}
\label{fig:qualitative_single}
\end{figure}

Fig.~\ref{fig:qualitative_single} illustrates the evolution of a RadFlux-BC sample through factor composition. The final $15/15$ prediction preserves the main hot-spot locations, boundary-induced gradients, and overall heat-diffusion trend. Residuals are concentrated near local heat-source neighborhoods and boundary high-gradient regions. This observation is consistent with the quantitative diagnostics: the composed representation retains the large-scale field structure, while the remaining errors mainly concern local details.

\section{Conclusion}
\label{sec:conclusion}

This study proposed the Tree-Structured Factor Composition Network (TFCN) for rapid thermal-field evaluation of modular spacecraft panels with varying component numbers and layouts. TFCN represents heat-source components and boundary segments as object-level factors and progressively combines their encoded representations through a shared composer before decoding the complete temperature field.

In both boundary-condition settings, TFCN achieved the lowest mean error for unseen component counts. In Temp-BC, TFCN-TP reduced the mean $K$-\ood\ \rmse\ from $12.23$ K for GINO to $4.21$ K, while in RadFlux-BC it reduced the mean from $28.05$ K to $18.62$ K. Qualitative field-level results further showed that TFCN preserved the main hot-spot locations and boundary-induced gradients. These results support the use of TFCN for rapidly pre-screening candidate spacecraft thermal configurations before high-fidelity numerical analysis and verification.

The present conclusions are restricted to two-dimensional steady-state thermal problems, $K$-\ood\ evaluation, and the factor and boundary types represented in the training data. The current implementation also uses a predefined factor capacity and a balanced binary composition structure. Future work will evaluate three-dimensional, transient, cross-geometry, and more strongly nonlinear problems, together with variable-capacity inputs and alternative composition mechanisms.

\section*{Conflict of Interest}

The authors declare that there are no conflicts of interest, whether financial or personal, that could have influenced the findings or interpretation of the research presented in this paper.

\section*{Data Availability}
The code can be obtained by contacting the corresponding author upon reasonable request.

\section*{Acknowledgments}

This work was supported by the Science and Technology Commission of Shanghai Municipality (STCSM) Key Technology R\&D Plan (No. 25DZ3100700) and the STCSM Shanghai Science and Technology Innovation Action Plan (No. 24DZ3101400).

\appendix
\setcounter{table}{0}

\section{Notation and model implementation details}
\label{app:notation_implementation}

\subsection{Principal notation}

Table~\ref{tab:principal_notation} summarizes the principal notation used in TFCN.

\begin{table}[H]
\centering
\caption{Summary of principal notation used in TFCN.}
\label{tab:principal_notation}
\footnotesize
\renewcommand{\arraystretch}{1.08}
\begin{tabularx}{0.96\linewidth}{@{}>{\centering\arraybackslash}p{0.23\linewidth}X@{}}
\toprule
Symbol & Meaning \\
\midrule
$\Omega_h$ & Discrete grid \\
$G$ & Coordinate condition \\
$\mathcal{C}$ & Heat-source components \\
$\mathcal{B}$ & Boundary segments \\
$\mathcal{F}$ & Factors \\
$f_i$ & Factor representation \\
$D,D_{\max}$ & Current/maximum depth \\
$Q=2^D$ & Bucket/leaf-token count, not size \\
$L_Q$ & Factor-slot capacity \\
$X_j$ & Bucket feature \\
$U_j$ & Encoded token \\
$M_{\phi}$ & Shared composer \\
$Z$ & Composed representation \\
$D_{\psi}$ & Temperature-field decoder \\
$T,\hat{T}$ & True/predicted physical fields \\
\bottomrule
\end{tabularx}
\end{table}

\subsection{Factor-slot padding and masking}
\label{app:factor_padding}

Specifically, the bucketing operator $\Pi_Q$ first obtains bucket index groups $\{\mathcal{P}_j\}_{j=1}^{Q}$. For each $\mathcal{P}_j$, the model keeps the SDF channel and value feature of each real factor, and pads empty factor slots when needed to obtain a fixed tensor shape. Let $L_Q$ denote the fixed factor-slot capacity of each bucket. For the source-factor input, the present implementation uses $L_Q=20$. After balanced assignment of $K$ source factors to $Q$ buckets, each bucket contains at most $\lceil K/Q\rceil$ real factors; hence the input is valid when $\lceil K/Q\rceil\leq L_Q$. Unused positions are filled with predefined padding values and excluded by the validity mask. Let the padded bucket index sequence be $\widetilde{\mathcal{P}}_j=(i_{j,1},\ldots,i_{j,L_Q})$, where an empty placeholder is denoted by $\varnothing$. Define $d_{\varnothing}(\mathbf{x})=d_{\mathrm{pad}}>\mathrm{diam}(\Omega)$ and $\mathbf{v}_{\varnothing}=\mathbf{0}$. The pre-encoding bucket feature and the corresponding encoded token are defined as:
\begin{equation}
\begin{aligned}
    m_{j,\ell}
    &=\mathbb{I}[i_{j,\ell}\neq\varnothing],\\
    X_j(\mathbf{x})
    &=
    \bigoplus_{\ell=1}^{L_Q}
    m_{j,\ell}
    \left(
    d_{i_{j,\ell}}(\mathbf{x}),
    \mathbf{v}_{i_{j,\ell}}\mathbf{1}_{\Omega_h}(\mathbf{x})
    \right),\\
    U_j
    &=E_{\theta}(X_j),\qquad
    U_j\in\mathbb{R}^{H\times W\times d_z}.
\end{aligned}
\end{equation}
Here $\bigoplus$ denotes channel-wise concatenation, $\mathbf{1}_{\Omega_h}$ broadcasts $\mathbf{v}_i$ over the grid, and $m_{j,\ell}\in\{0,1\}$ is the factor-slot validity mask. The validity mask zeros padded factor slots before bucket encoding. This input padding is applied independently of the $4\times4$ spatial-window partition used by the composer.

\subsection{Network structure and scale}

Tables~\ref{tab:appendix_unet_architecture}--\ref{tab:appendix_tfcn_architecture} report the detailed network structures and parameter scales of the compared models.

\begin{table}[H]
\centering
\scriptsize
\renewcommand{\arraystretch}{1.05}
\caption{Detailed UNet architecture.}
\label{tab:appendix_unet_architecture}
\vspace{0.3em}
\begin{tabular}{>{\raggedright\arraybackslash}p{0.58\linewidth}>{\raggedright\arraybackslash}p{0.32\linewidth}}
\toprule
Network type & Output size \\
\midrule
Input spatial fields & $N \times 64 \times 64 \times 42$ \\
Encoder blocks $\times 4$ & $N \times \{32,64,128,256\} \times \{64,32,16,8\}^{2}$ \\
Bottleneck & $N \times 512 \times 4 \times 4$ \\
Up-conv + skip decoder blocks $\times 4$ & $N \times \{256,128,64,32\} \times \{8,16,32,64\}^{2}$ \\
Conv2dResBlock + $3\times3$ Conv & $N \times 1 \times 64 \times 64$ \\
Output temperature field & $N \times 64 \times 64 \times 1$ \\
\midrule
\multicolumn{2}{p{0.94\linewidth}}{\textbf{Total params:} $8{,}566{,}017$} \\
\multicolumn{2}{p{0.94\linewidth}}{\textbf{Params size:} $32.68$ MB, FP32} \\
\multicolumn{2}{p{0.94\linewidth}}{\textbf{Forward pass size:} $\approx 18.53$ MB} \\
\bottomrule
\end{tabular}
\end{table}

\begin{table}[H]
\centering
\scriptsize
\renewcommand{\arraystretch}{1.05}
\caption{Detailed DeepONet architecture.}
\label{tab:appendix_deeponet_architecture}
\vspace{0.3em}
\begin{tabular}{>{\raggedright\arraybackslash}p{0.58\linewidth}>{\raggedright\arraybackslash}p{0.32\linewidth}}
\toprule
Network type & Output size \\
\midrule
Branch input fields & $N \times 64 \times 64 \times 39$ \\
Flatten branch input & $N \times 159744$ \\
Branch MLP: Linear + GELU $\times 2$ + Linear & $N \times 64$ \\
Trunk coordinate input & $N \times 64 \times 64 \times 2$ \\
Trunk MLP: Linear + GELU $\times 2$ + Linear & $N \times 64 \times 64 \times 64$ \\
Branch broadcast + branch--trunk product & $N \times 64 \times 64 \times 64$ \\
Summation head & $N \times 64 \times 64 \times 1$ \\
Output temperature field & $N \times 64 \times 64 \times 1$ \\
\midrule
\multicolumn{2}{p{0.94\linewidth}}{\textbf{Total params:} $10{,}240{,}512$} \\
\multicolumn{2}{p{0.94\linewidth}}{\textbf{Params size:} $39.06$ MB, FP32} \\
\multicolumn{2}{p{0.94\linewidth}}{\textbf{Forward pass size:} $\approx 8.67$ MB} \\
\bottomrule
\end{tabular}
\end{table}

\begin{table}[H]
\centering
\scriptsize
\renewcommand{\arraystretch}{1.05}
\caption{Detailed GINO architecture.}
\label{tab:appendix_gino_architecture}
\vspace{0.3em}
\begin{tabular}{>{\raggedright\arraybackslash}p{0.40\linewidth}>{\raggedright\arraybackslash}p{0.52\linewidth}}
\toprule
Network stage & Output size and configuration \\
\midrule
Input spatial fields & $N \times 64 \times 64 \times 9$ \\
Normalized spatial coordinates & $1 \times 4096 \times 2$ (shared geometry) \\
Input GNO integral transform & $N \times 16 \times 16 \times 9$; radius $0.11$; nonlinear MLP $265\rightarrow64\rightarrow64\rightarrow9$ \\
Channel lifting & $N \times 48 \times 16 \times 16$; $9\rightarrow96\rightarrow48$ \\
FNO blocks $\times 4$ \citep{li2021fno} & $N \times 48 \times 16 \times 16$; nominal modes $(16,16)$; spectral weights $48 \times 48 \times 16 \times 9$ \\
Channel MLP in each FNO block & $N \times 48 \times 16 \times 16$; $48\rightarrow24\rightarrow48$ \\
Output GNO integral transform & $N \times 4096 \times 48$; radius $0.11$; nonlinear MLP $304\rightarrow128\rightarrow64\rightarrow48$ \\
Output projection & $N \times 4096 \times 1$; $48\rightarrow192\rightarrow1$ \\
Output temperature field & $N \times 64 \times 64 \times 1$ \\
\midrule
\multicolumn{2}{p{0.94\linewidth}}{\textbf{Total params:} $1{,}433{,}610$} \\
\multicolumn{2}{p{0.94\linewidth}}{\textbf{Parameter storage:} $\approx 10.53$ MiB (FP32/complex64)} \\
\multicolumn{2}{p{0.94\linewidth}}{\textbf{Peak forward memory:} $\approx 2.48$ GiB (batch size 1, FP32 benchmark)} \\
\bottomrule
\end{tabular}
\end{table}

\begin{table}[H]
\caption{Detailed TFCN architecture.}
\label{tab:appendix_tfcn_architecture}
\centering
\scriptsize
\renewcommand{\arraystretch}{1.05}
\begin{tabular}{>{\raggedright\arraybackslash}p{0.58\linewidth}>{\raggedright\arraybackslash}p{0.35\linewidth}}
\toprule
Network type & Output size \\
\midrule
\multicolumn{2}{l}{\textbf{Inputs}} \\
Input: geometry grid & $N \times 64 \times 64 \times 2$ \\
Input: component factors & SDF $N \times 64 \times 64 \times 20$; value $N \times 20$ \\
Input: boundary factors & SDF $N \times 64 \times 64 \times 2$; value $N \times 2$ \\
\midrule
\multicolumn{2}{l}{\textbf{Encoder}} \\
Bucket features and encoder tokens & $4 \times (N \times 64 \times 64 \times 32)$ \\
\midrule
\multicolumn{2}{l}{\textbf{Composer}} \\
Pairwise tree composer, up to 2 levels & $N \times 64 \times 64 \times 32$ \\
\midrule
\multicolumn{2}{l}{\textbf{Decoder}} \\
Dual-head embedding, $32\rightarrow48$ & $N \times 4096 \times 48$ \\
Transformer self-attention layers $\times2$ \citep{vaswani2017attention} & $N \times 4096 \times 48$ \\
Fourier regressor + linear head & $N \times 64 \times 64 \times 1$ \\
Output temperature field & $N \times 64 \times 64 \times 1$ \\
\midrule
\multicolumn{2}{p{0.94\linewidth}}{\textbf{Total params:} $393{,}170$} \\
\multicolumn{2}{p{0.94\linewidth}}{\textbf{Params size:} $1.50$ MB, FP32} \\
\multicolumn{2}{p{0.94\linewidth}}{\textbf{Forward pass size:} $\approx 56.27$ MB} \\
\bottomrule
\end{tabular}
\end{table}

\section{Data and training configuration}
\label{app:data_training}

\subsection{Data distribution diagnostics}

\begin{figure}[H]
\centering
\begin{minipage}{0.75\linewidth}
\centering

\begin{minipage}{0.49\linewidth}
\centering
\includegraphics[height=0.82\linewidth]{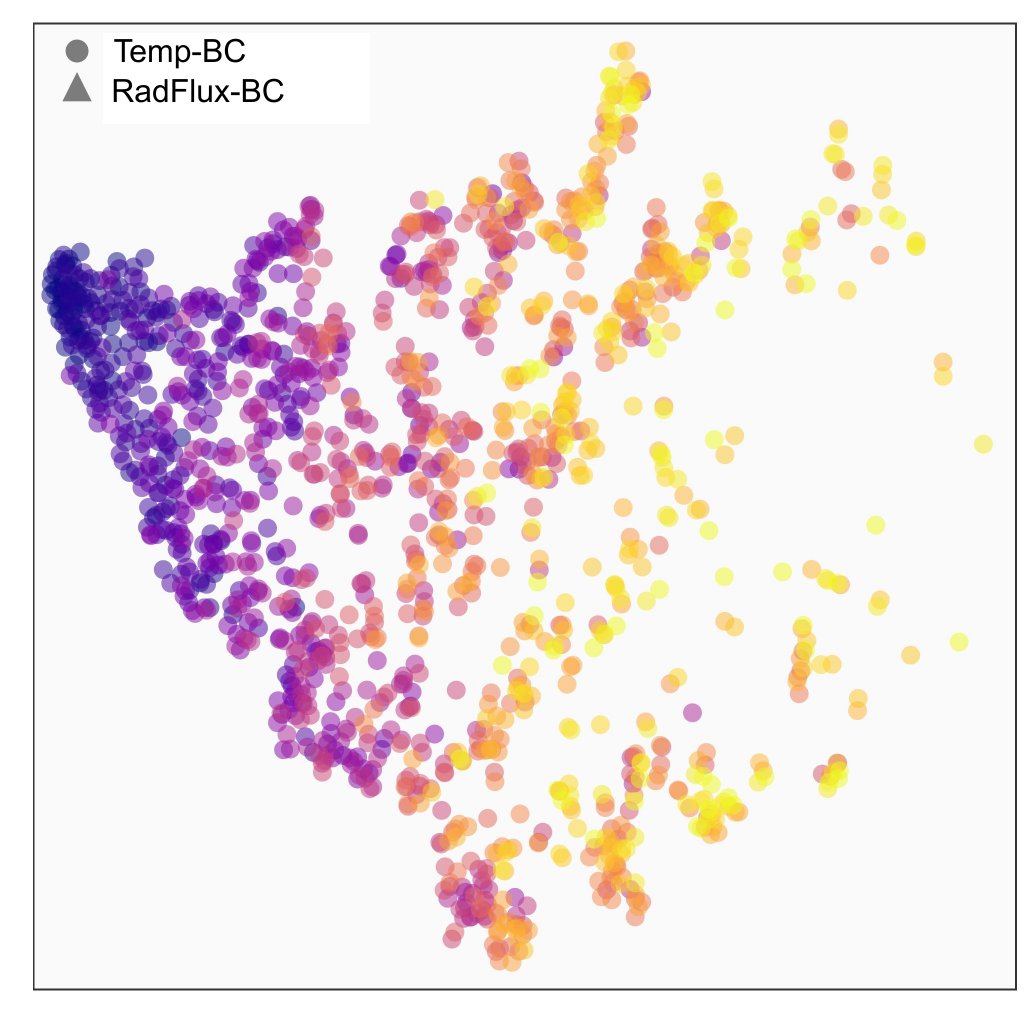}
\par\vspace{0.1em}{\footnotesize (a) Temp-BC distribution}
\end{minipage}
\hfill
\begin{minipage}{0.49\linewidth}
\centering
\includegraphics[height=0.82\linewidth]{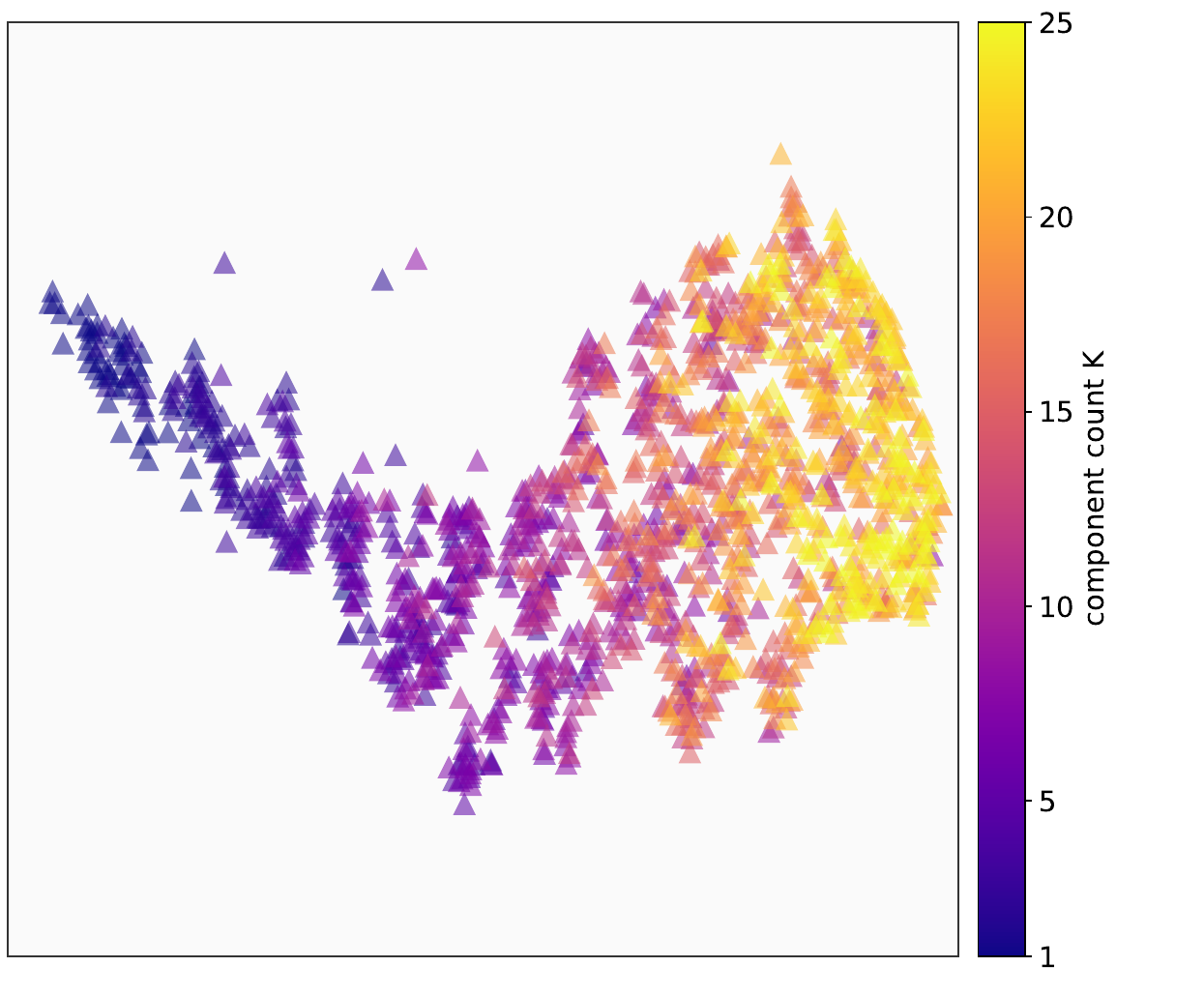}
\par\vspace{0.1em}{\footnotesize (b) RadFlux-BC distribution}
\end{minipage}

\end{minipage}
\caption{Temperature-field sample distribution diagnostics.}
\label{fig:dataset_tsne}
\end{figure}

Fig.~\ref{fig:dataset_tsne} shows the low-dimensional distributions of temperature-field samples in the Temp-BC and RadFlux-BC scenarios.

\subsection{Training and evaluation configuration}

All models were trained with AdamW \citep{loshchilov2019adamw} for 1500 epochs using a batch size of 32. The learning rate was linearly warmed up to $8\times10^{-4}$ over 60 epochs and then reduced to $1\times10^{-5}$ by cosine decay. Weight decay was $1\times10^{-4}$, and the gradient norm was clipped at 1.0. Early stopping was not used. Checkpoints were selected only by error on a fixed ID validation set; OOD results were not used. Table~\ref{tab:training_configuration} summarizes the training and evaluation settings.

\begin{table}[H]
\centering
\small
\caption{Training and evaluation configuration.}
\label{tab:training_configuration}
\renewcommand{\arraystretch}{1.08}
\begin{tabular}{ll}
\toprule
Item & Setting \\
\midrule
Optimizer & AdamW \\
Training epochs & 1500 \\
Batch size & 32 \\
Learning rate & $8\times10^{-4}$ peak; $1\times10^{-5}$ final \\
Warm-up and decay & 60 epochs, linear; cosine decay \\
Weight decay & $1\times10^{-4}$ \\
Gradient clipping & 1.0 \\
Early stopping & Not used \\
Checkpoint selection & Fixed ID validation error \\
OOD data used for selection & No \\
\bottomrule
\end{tabular}
\end{table}

\section{Supplementary evaluation metrics and efficiency}
\label{app:supplementary_evaluation}

\subsection{Physics-oriented thermal indicators}

Tables~\ref{tab:thermal_indicator_definitions} and~\ref{tab:thermal_indicator_results} define the physics-oriented indicators and report the corresponding ID/OOD results together with supplementary RMSE statistics.

\begin{table}[H]
\centering
\caption{Definitions of the physics-oriented thermal indicators.}
\label{tab:thermal_indicator_definitions}
\footnotesize
\setlength{\tabcolsep}{4pt}
\renewcommand{\arraystretch}{1.12}
\begin{tabularx}{\linewidth}{@{}>{\centering\arraybackslash}p{0.18\linewidth}>{\raggedright\arraybackslash}p{0.28\linewidth}>{\centering\arraybackslash}X@{}}
\toprule
Symbol & Physical meaning & Definition \\
\midrule
$E_{T_{\max}}^{\mathrm{mean}}$
& Mean peak-temperature deviation (K)
&
$\displaystyle
\frac{1}{N}\sum_{i=1}^{N}
\left|\hat{T}_{\max,i}-T_{\max,i}\right|
$
\\[6pt]
$E_{T_{\max}}^{\mathrm{P95}}$
& Weighted P95 peak-temperature deviation (K)
&
$\displaystyle
Q_{0.95}^{(w)}\!\left(
\left\{
\left|\hat{T}_{\max,i}-T_{\max,i}\right|
\right\}_{i=1}^{N}
\right)
$
\\[6pt]
$E_{\mathrm{hot}}^{\mathrm{rel}}$
& Mean relative hot-spot location deviation
&
$\displaystyle
\frac{1}{N}\sum_{i=1}^{N}
\frac{
\left\|
\hat{\boldsymbol{x}}_{\max,i}
-
\boldsymbol{x}_{\max,i}
\right\|_{2}
}{L_{\Omega}}
$
\\[6pt]
$\mathrm{RMSE}_{B}$
& Boundary-band temperature RMSE (K)
&
$\displaystyle
\left[
\frac{
\sum_{i=1}^{N}
\sum_{\boldsymbol{x}\in\Omega_{B,i}}
\left(
\hat{T}_{i}(\boldsymbol{x})-T_{i}(\boldsymbol{x})
\right)^2
}{
\sum_{i=1}^{N}|\Omega_{B,i}|
}
\right]^{1/2}
$
\\
\bottomrule
\end{tabularx}

\vspace{2pt}
\begin{minipage}{\linewidth}
\footnotesize
$Q_{0.95}^{(w)}$ denotes the weighted 95th percentile computed using the stratified-sampling weights. $L_{\Omega}$ is the diagonal length of the computational domain. $\Omega_{B,i}$ denotes the one-grid-wide boundary band extracted independently from each boundary-SDF channel.
\end{minipage}
\end{table}

\begin{table}[H]
\centering
\caption{Physics-oriented thermal indicators and supplementary RMSE statistics under ID/OOD evaluation.}
\label{tab:thermal_indicator_results}
\small
\setlength{\tabcolsep}{8pt}
\renewcommand{\arraystretch}{1.10}
\begin{tabularx}{\linewidth}{@{}Xccc@{}}
\toprule
Indicator $\downarrow$
& GINO
& TFCN-SP
& TFCN-TP \\
\midrule
\multicolumn{4}{@{}l}{\textbf{Temperature BC}} \\
$E_{T_{\max}}^{\mathrm{mean}}$ (K)
& 3.14 / 8.03
& 2.12 / 7.67
& \textbf{1.34 / 2.92} \\
$E_{T_{\max}}^{\mathrm{P95}}$ (K)
& 9.09 / 26.41
& 6.83 / 28.44
& \textbf{4.13 / 8.91} \\
$E_{\mathrm{hot}}^{\mathrm{rel}}$
& 0.118 / 0.078
& 0.088 / 0.090
& \textbf{0.058 / 0.068} \\
$\mathrm{RMSE}_{B}$ (K)
& 5.16 / 14.08
& 1.81 / 5.82
& \textbf{1.22 / 3.70} \\
$\mathrm{RMSE}$ (worst case, K)
& 22.29 / 77.00
& 56.22 / 82.53
& \textbf{8.35 / 45.80} \\
$\mathrm{RMSE}$ ($K=25$, K)
& 31.35
& 17.67
& \textbf{5.64} \\
\midrule
\multicolumn{4}{@{}l}{\textbf{Radiative-flux BC}} \\
$E_{T_{\max}}^{\mathrm{mean}}$ (K)
& 7.73 / 15.49
& 4.22 / 13.03
& \textbf{2.47 / 8.24} \\
$E_{T_{\max}}^{\mathrm{P95}}$ (K)
& 21.15 / 48.88
& 12.28 / 38.92
& \textbf{7.33 / 25.87} \\
$E_{\mathrm{hot}}^{\mathrm{rel}}$
& 0.138 / 0.114
& 0.107 / 0.110
& \textbf{0.081 / 0.094} \\
$\mathrm{RMSE}_{B}$ (K)
& 6.85 / 14.62
& 3.00 / 8.81
& \textbf{1.79 / 6.41} \\
$\mathrm{RMSE}$ (worst case, K)
& 66.91 / 127.58
& 92.32 / 107.41
& \textbf{33.63 / 83.85} \\
$\mathrm{RMSE}$ ($K=25$, K)
& 42.20
& 28.30
& \textbf{16.50} \\
\bottomrule
\end{tabularx}

\vspace{2pt}
\begin{minipage}{\linewidth}
\footnotesize
The first four rows are the physics-oriented thermal indicators defined in Table~\ref{tab:thermal_indicator_definitions}, whereas the last two rows report supplementary full-field RMSE statistics. Values are presented as ID/OOD, except for $\mathrm{RMSE}$ ($K=25$), which is evaluated only on the $K=25$ OOD subset.
\end{minipage}
\end{table}

\subsection{Computational efficiency and workflow role}

Table~\ref{tab:computational_efficiency} compares the reference solver with the trained surrogate. Offline training is required once, after which TFCN reduces evaluation of 5,000 candidates from hours or days to seconds. TFCN is used for preliminary screening, and selected configurations are verified with FEniCSx \citep{baratta2023dolfinx}.

\begin{table}[H]
\centering
\caption{Computational efficiency under two boundary-condition settings.}
\label{tab:computational_efficiency}
\footnotesize
\renewcommand{\arraystretch}{1.15}
\begin{tabularx}{0.98\linewidth}{@{}l>{\raggedright\arraybackslash}p{0.18\linewidth}>{\centering\arraybackslash}p{0.12\linewidth}>{\centering\arraybackslash}X>{\centering\arraybackslash}X@{}}
\toprule
Method & Boundary condition & Single config. &
5,000 sequential & 5,000 parallel/batched \\
\midrule
FEniCSx & Non-radiative & 288 s & 16.7 d & 6.25 h$^{*}$ \\
FEniCSx & Radiative     & 320 s & 18.5 d & 6.94 h$^{*}$ \\
TFCN    & Non-radiative & 7 ms  & 35 s$^{\dagger}$ & 1.8 s$^{\ddagger}$ \\
TFCN    & Radiative     & 7 ms  & 35 s$^{\dagger}$ & 1.8 s$^{\ddagger}$ \\
\bottomrule
\end{tabularx}

\vspace{2pt}
\begin{minipage}{0.96\linewidth}
\footnotesize
\raggedright
\textit{Note:} $^{*}$Ideal 64-way parallelism;
$^{\dagger}$estimated from single-configuration latency;
$^{\ddagger}$batched inference at 2,800 samples/s.
\end{minipage}
\end{table}

\section{Supplementary field-level prediction samples}
\label{app:field_samples}

\begin{figure}[htbp]
\centering
\includegraphics[width=0.85\linewidth]{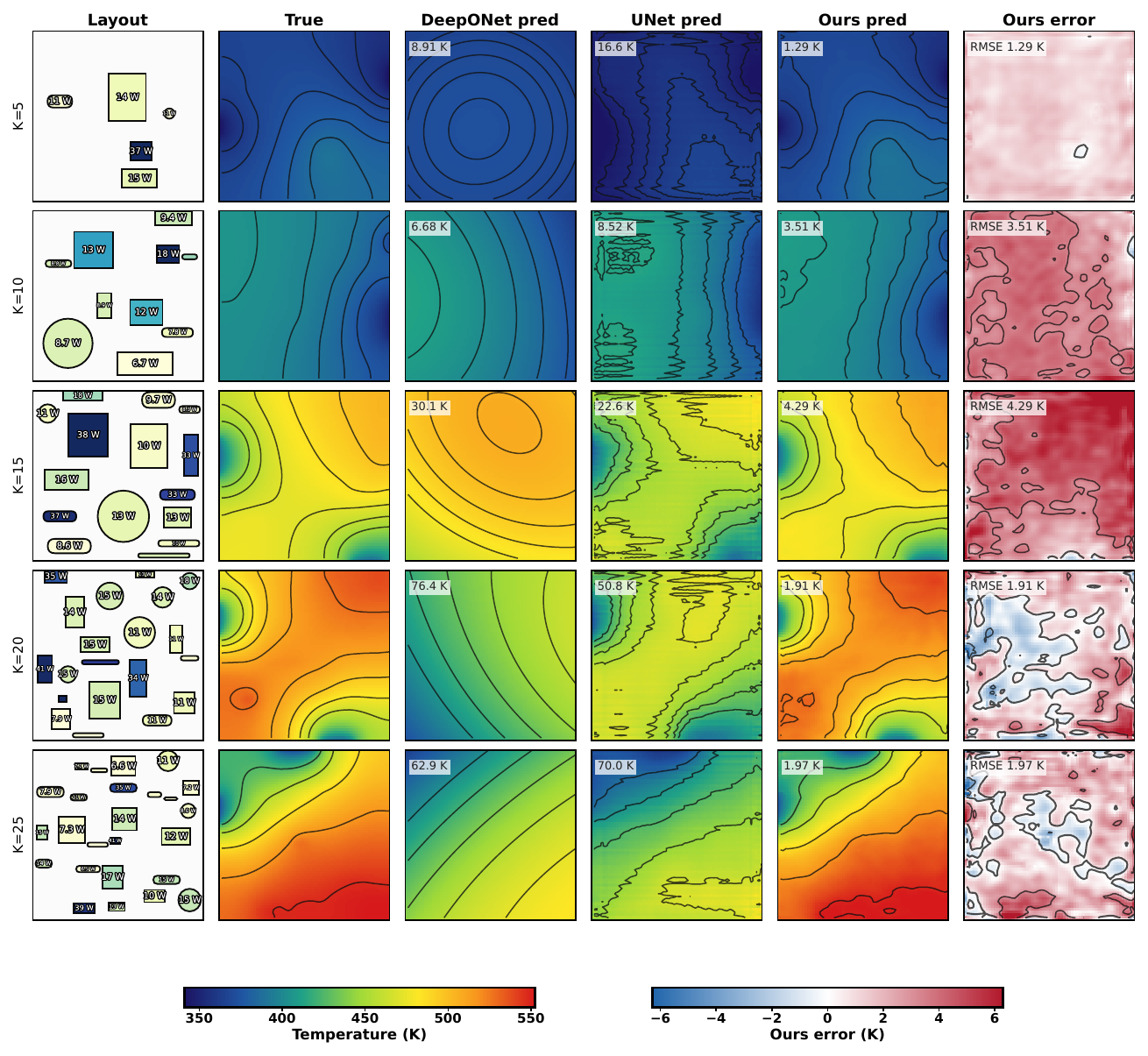}
\caption{RadFlux-BC temperature-field prediction.}
\label{fig:qualitative_multi}
\end{figure}

\begin{figure}[H]
\centering
\includegraphics[width=0.85\linewidth]{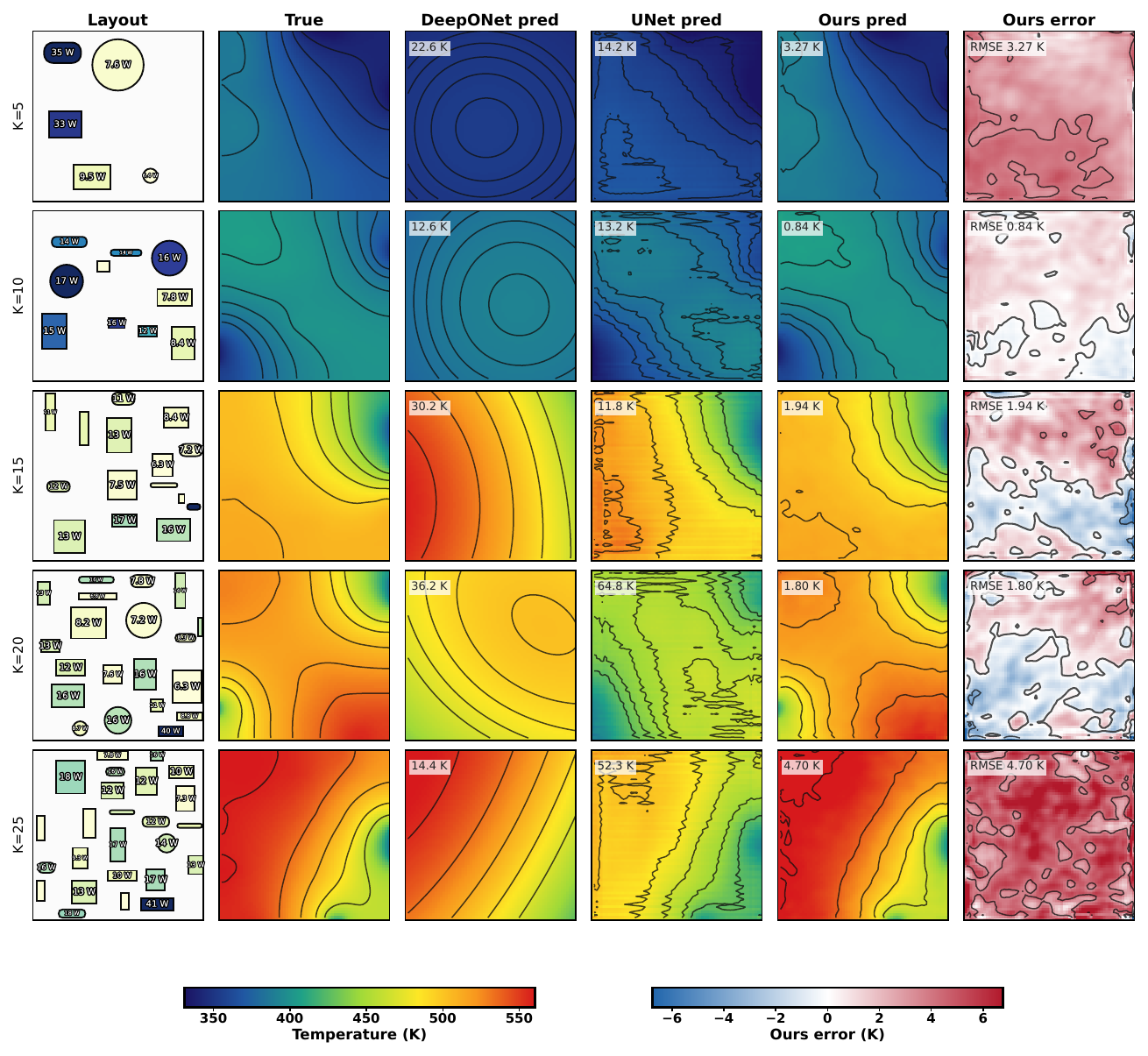}
\caption{Supplementary RadFlux-BC temperature-field prediction samples.}
\label{fig:qualitative_multi_appendix}
\end{figure}

Figs.~\ref{fig:qualitative_multi} and~\ref{fig:qualitative_multi_appendix} provide two batches of RadFlux-BC temperature-field prediction samples. They complement the quantitative results with field-level comparisons of the reference and predicted temperature distributions.

\bibliographystyle{elsarticle-num}
\bibliography{references}

\end{document}